\documentclass[12pt]{article}
\usepackage{putex}
\usepackage{graphicx}

\begin{document}

\preprint{PUPT-2270}

\institution{PU}{Joseph Henry Laboratories, Princeton University, Princeton, NJ 08544}

\title{The gravity dual of a $p$-wave superconductor}

\authors{Steven S. Gubser and Silviu S. Pufu}

\abstract{We construct black hole solutions to the Yang-Mills equations in an $AdS_4$-Schwarzschild background which exhibit superconductivity.  What makes these backgrounds $p$-wave superconductors is that the order parameter is a vector, and the conductivities are strongly anisotropic in a manner that is suggestive of a gap with nodes.  The low-lying excitations of the normal state have a relaxation time which grows rapidly as the temperature decreases, consistent with the absence of impurity scattering.  A numerical exploration of quasinormal modes close to the transition temperature suggests that $p$-wave backgrounds are stable against perturbations analogous to turning on a $p+ip$ gap, whereas $p+ip$-wave configurations are unstable against turning into pure $p$-wave backgrounds.}

\date{May 2008}

\maketitle

\tableofcontents

\section{Introduction}

In \cite{Gubser:2005ih,Gubser:2008px,Hartnoll:2008vx,Gubser:2008zu}, three variants of a mechanism were proposed through which black holes can spontaneously break an abelian gauge symmetry at finite temperature, leading to a form of superconductivity.  The symmetry breaking occurs through the formation of a superconducting condensate that floats above the horizon.  To make the mechanism work, one needs a charged matter field whose quanta form the superconducting layer, some interaction which keeps it from falling into the black hole, and some feature which prevents it from escaping to infinity.  In the variants considered so far, the charged field is bosonic.  Charged fermions might also work, but then one has to establish some pairing mechanism before condensation can occur.

The purpose of this paper is to provide another example of superconducting black holes, closely related to the one in \cite{Gubser:2008zu}, and to argue that it exhibits a $p$-wave gap.  All our results will be based on classical solutions to field equations of Einstein-Yang-Mills theory with a negative cosmological constant:
 \eqn{EYMS}{
  S = {1 \over 2\kappa^2} \int d^4 x \, \left[
    R - {1 \over 4} (F_{\mu\nu}^a)^2 + {6 \over L^2} \right] \,,
 }
where $F_{\mu\nu}^a$ is the field strength of an $SU(2)$ gauge field.  As in \cite{Gubser:2008zu}, the plan is to regard a $U(1)$ subgroup as the gauge group of electromagnetism,\footnote{More precisely, the boundary theory has a global $SU(2)$ symmetry, and adding electromagnetism means weakly gauging this $U(1)$ in the boundary theory.  By contrast, the gauging of the full $SU(2)$ symmetry in the gravity theory encodes aspects of the $SU(2)$ current algebra dynamics in the boundary theory.} and to persuade the off-diagonal gauge bosons, charged under this $U(1)$, to condense outside the horizon.  Our conventions on metric signature and field normalizations are the same as in \cite{Gubser:2008zu}.  The action \eno{EYMS} is almost completely dictated at the two-derivative level by local diffeomorphism symmetry and gauge symmetry, and it can be embedded in M-theory.\footnote{Embedding \eno{EYMS} in string theory or M-theory involves a specific choice of the gauge coupling $g$, and it has not been shown that any of the three variants \cite{Gubser:2005ih,Gubser:2008px,Gubser:2008zu} of superconducting black holes occurs in M-theory.  Also, M-theory constructions usually involve scalars that couple non-renormalizably to the gauge field and may be an important ingredient in constructing symmetry breaking solutions.}  This endows its solutions with an interest independent of potential applications to superconductivity.  Indeed, the solutions to be considered are loosely related to those of \cite{Bartnik:1988am,Bizon:1990sr}, from which a significant literature has sprung, reviewed for example in \cite{Volkov:1998cc,Winstanley:2008ac}.

Section~\ref{OVERVIEW} is devoted to a conceptual overview of black holes which superconduct through the mechanism described above.  In section~\ref{BACKGROUND} we describe the background solutions of interest.  In section~\ref{PERTURBATIONS} we study the electromagnetic response, along the lines of \cite{Hartnoll:2008vx}.  We find a frequency-dependent conductivity which depends strongly on the polarization of the applied electric field.  The low-frequency behavior is suggestive of quasi-particle excitations whose dissipative mechanisms are entirely due to finite-temperature effects.

In section~\ref{STABILITY} we provide numerical evidence that the $p$-wave backgrounds are stable against small perturbations that turn on a $p+ip$ gap.  In section~\ref{PPLUSIP} we provide numerical evidence that the $p+ip$-wave backgrounds of \cite{Gubser:2008px} are unstable against small perturbations that turn them into the $p$-wave backgrounds described in section~\ref{BACKGROUND}.  Our numerical explorations are far from covering the full range of parameters, but the simplest scenario consistent with them is that $p+ip$-wave backgrounds are always unstable, and that $p$-wave backgrounds represent the thermodynamically preferred phase for $T$ less than a critical temperature $T_c$.

\section{Conceptual overview}
\label{OVERVIEW}

\begin{figure}
  \centerline{\includegraphics[width=6in]{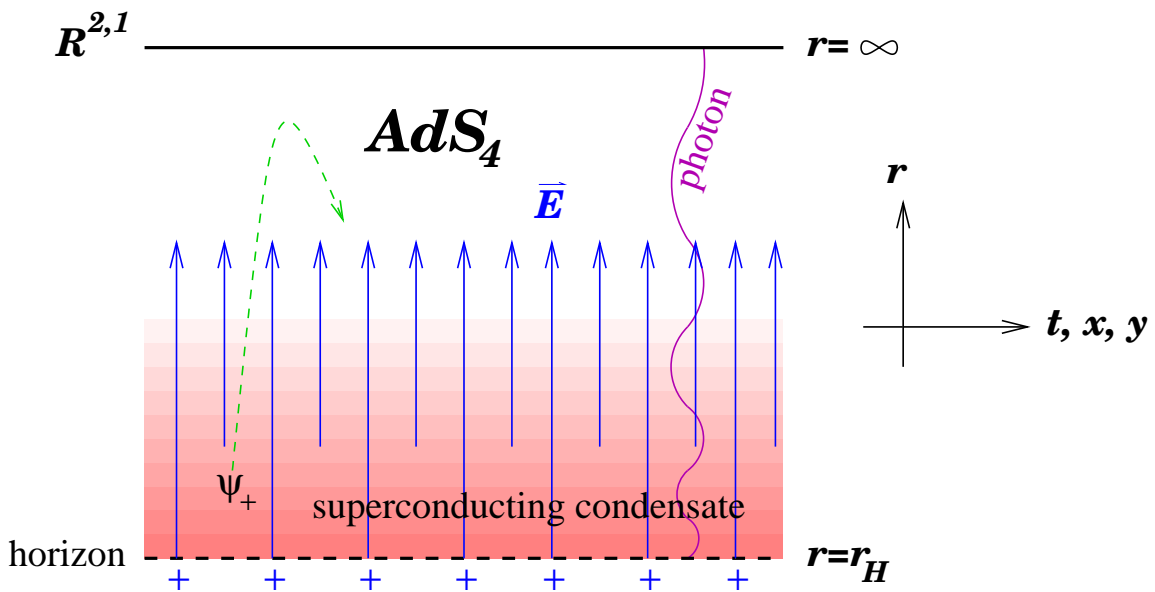}}
  \caption{A superconducting condensate floats above a black hole horizon because of a balance of gravitational and electrostatic forces.  The condensate carries a finite fraction of the total charge density, so there is more electric flux above the condensate than there is right at the horizon.  A massive charged particle, labeled $\psi_+$, may be driven upward by the electrostatic force, but because of the warped geometry of $AdS_4$, its trajectory cannot reach the boundary.  So $\psi_+$ must participate in the condensate if it doesn't fall into the horizon.  The frequency-dependent conductivity can be found by calculating an on-shell amplitude for a photon propagating straight down into the geometry.}\label{BALANCE}
 \end{figure}

In the anti-de Sitter examples of \cite{Gubser:2008px,Hartnoll:2008vx,Gubser:2008zu}, the superconducting layer floats above the horizon because the horizon is also charged.  Electrostatic repulsion overcomes the gravitational attraction that ordinarily would suck the superconducting layer into the horizon.  If the spacetime were asymptotically flat, then (barring some special interactions such as considered in \cite{Gubser:2005ih}) one expects that electrostatic repulsion would cause the superconducting layer to be blown off to infinity.  But asymptotically anti-de Sitter geometries prevent this.  Massive particles, no matter how strongly repelled from a horizon, cannot reach the boundary of anti-de Sitter space.  So they instead condense near the horizon, where ``near'' means that the field profile is normalizable, carrying finite charge if the horizon is finite, or finite charge density if the horizon has infinite extent.  An analogy with the classic two-fluid model of superconductors is possible: the charged horizon describes the normal component, and the condensate above it is the superconducting component.  See figure~\ref{BALANCE}.  In this analogy, it is important to recall that in the gauge-string duality \cite{Maldacena:1997re,Gubser:1998bc,Witten:1998qj}, the extra dimension $r$ is not an additional flat dimension transverse to the sample; instead, it is a way of organizing energy scales in the dual field theory, which is strictly $2+1$-dimensional and non-gravitational.  Thus, although the condensate is ``above'' the horizon in the gravity picture, it interpenetrates the normal state in the field theory picture.

In \cite{Gubser:2008zu}, where the bulk geometry is based on Einstein-Yang-Mills theory, it was shown that there is a second order transition, with mean field theory exponents, between a non-superconducting state at high temperatures, where all the charge is in the normal component, and a superconducting state at low temperatures.  A similar transition probably occurs for black hole geometries based on gravity coupled to the Abelian Higgs model \cite{Hartnoll:2008vx,Gubser:2008zu}, but this has not been demonstrated conclusively.  What is clear, both for Einstein-Yang-Mills and for the Abelian Higgs model, is that the moduli space of black hole solutions includes Reissner-Nordstr\"om anti-de Sitter black holes (hereafter RNAdS), which describe the normal state,\footnote{It has been suggested \cite{Hartnoll:2007ih,Hartnoll:2007ip} that RNAdS black holes are dual to a close analog of the pseudogap state of high $T_c$ materials.  In the context of our constructions, this does not seem quite right, because the fraction of charge in the condensate goes to zero near $T_c$, scaling as $T_c-T$, whereas the transition from superconductivity to the pseudogap state appears to take place while this fraction is finite and non-zero.} joining continuously onto a branch of symmetry-breaking solutions.  The simplest argument supporting this picture is based on studying linear perturbations of the charged field around an RNAdS solution.  They obey an equation of the form
 \eqn{ChargedPerturb}{
  (\square - m_{\rm eff}^2) \psi = 0 \,,
 }
where $\square$ is an appropriate covariant wave operator and
 \eqn{meff}{
  m_{\rm eff}^2 = m^2 + g^{tt} q^2 \Phi^2 \,.
 }
Here $q$ is the charge of a quantum of the charged bosonic field $\psi$: in the Abelian Higgs case, it is a complex scalar, while in the Einstein-Yang-Mills case, it is a complex combination of non-abelian gauge bosons.  $\Phi$ is the electrostatic potential, which vanishes at the horizon but grows quickly outside it if the electric field is strong.  The metric component $g^{tt}$ is negative in the conventions we use, and it diverges to $-\infty$ at the horizon, so \eno{meff} implies that $\psi$ is tachyonic near the horizon if $q$ is big enough and $m$ is small enough, provided also that the horizon carries sufficient charge.\footnote{Actually, the most commonly considered cases have $m^2 < 0$ in the case of scalars, or $m=0$ for non-abelian gauge bosons.  The argument about massive particles' trajectories never reaching the boundary of anti-de Sitter space then no longer holds up, but it is replaced by standard notions of boundary conditions in anti-de Sitter space which again lead to normalizable condensates.}  It is a matter of calculation to determine when \eno{ChargedPerturb} admits a static solution.  When it does, one may reasonably assume that it signifies the joining of a branch of symmetry breaking solutions onto the RNAdS solutions.  And one may calculate a critical temperature $T_c$ where the joining occurs.  It does not necessarily follow that $T_c$ is the temperature of a second order phase transition: it could be that the solutions which only slightly break the symmetry are thermodynamically disfavored, and that a first order transition to solutions with finite symmetry breaking occurs at a different temperature.  Explicit calculations of the free energy, as in \cite{Gubser:2008zu}, are necessary in order to determine the phase diagram.  But even the calculations of \cite{Gubser:2008zu} are not enough: one must also ask whether a solution is stable under small perturbations.  At least for a certain range of parameters, we will show in section~\ref{PPLUSIP} that the solutions of \cite{Gubser:2008zu} are unstable against a perturbation that seems likely to turn them into $p$-wave solutions of the form described in section~\ref{BACKGROUND}.  We have not yet found an unstable perturbation of the $p$-wave solutions, or of the $s$-wave solutions described in \cite{Gubser:2008px,Hartnoll:2008vx}.

In \cite{Hartnoll:2008vx}, the response of superconducting black holes to electromagnetic probes was studied.  The black holes in question were constructed along the lines of the proposal of \cite{Gubser:2008px}, by coupling the Abelian Higgs model to gravity, but a simplification was achieved by assuming that $q$ is large.  Provided $T$ is not too small relative to the charge density of the horizon, this implies that the back-reaction of both the gauge field and the charged scalar on the metric can be neglected.  A similar limit was considered in \cite{Gubser:2008zu} for the Einstein-Yang-Mills case, where by taking the gauge coupling large and avoiding the regime of very small temperatures, one may similarly neglect back-reaction of the matter fields on the metric.  A related limit of the proposal of \cite{Gubser:2008px} was studied in \cite{Albash:2008eh} in the presence of magnetic field, leading to partial localization of the condensate in one of the spatial directions on the boundary.  Also in the no-back-reaction limit, the effects of critical magnetic fields and vortices have been studied in \cite{Nakano:2008xc,Wen:2008pb}.

The quantity of primary interest in understanding the electromagnetic response is the conductivity,
 \eqn{sigmaDef}{
  \sigma_{ij}(\omega) = {i \over \omega} G^R_{ij}(\omega,0) \,,
 }
where
 \eqn{GRdef}{
  G^R_{mn}(\omega,\vec{k}) = -i \int d^3 x \,
    e^{i\omega t - i \vec{k} \cdot \vec{x}}
    \theta(t) \langle [J_m(t,\vec{x}),J_n(0,0)] \rangle
 }
is the retarded Green's function of the electromagnetic current $J_m$.\footnote{Our convention for indices is that $i$ and $j$ run over the spatial directions $x$ and $y$; $m$ and $n$ run over the boundary directions $t$, $x$, and $y$; and $\mu$ and $\nu$ run over the bulk directions $t$, $x$, $y$, and $r$.  We will later introduce adjoint indices $a$ and $b$ for the non-abelian gauge group $SU(2)$.}  The angle brackets in \eqref{GRdef} denote expectation values at finite temperature, namely
 \eqn{FiniteTempEV}{
  \langle {\cal A} \rangle \equiv {1\over Z} \tr e^{-\beta H} {\cal A}  \qquad Z \equiv \tr e^{-\beta H}
 }
for any operator ${\cal A}$.  The hermitian part of $\sigma_{ij}$ is dissipative, while the anti-hermitian part is reactive.\footnote{In the theory of AC circuits it is standard to consider the complex power $S = \int d^2 x \, E_i^* j_i = \int d^2 x\, E_i^* \sigma_{ij} E_j$, whose real and imaginary parts are the real and reactive powers, respectively.  The real power $P$ can therefore be expressed in terms of the hermitian part of $\sigma_{ij}$ through $P = \int d^2 x\, E_i^* {1\over 2} (\sigma_{ij} + \sigma_{ji}^*) E_j$, while the reactive power $Q = \int d^2 x\, E_i^* {1\over 2i} (\sigma_{ij} - \sigma_{ji}^*) E_j$ corresponds to the anti-hermitian part.}  According to a spectral decomposition, the hermitian part of $\sigma_{ij}$ should be positive semi-definite.  To see this, first note that the spacetime dependence of the hermitian operators $J_i(t, \vec{x})$ is found through
 \eqn{JSpaceTimeDep}{
  J_i(t, \vec{x}) = e^{i H t - i \vec{P} \cdot \vec{x}} J_i(0, 0) e^{-i H t + i \vec{P} \cdot \vec{x}} \,.
 }
Introducing a complete set of states between the two operators in \eqref{GRdef} and integrating over $t$ and $\vec{x}$ one obtains
 \eqn{GijDecomp}{
  G_{ij}^R(\omega, 0) = {1\over Z} \sum_{n, m} (2\pi)^2 \delta^{(2)}(\vec{P}_{nm}) J_i^{nm} J_j^{mn}
    {e^{-\beta E_n} - e^{-\beta E_m} \over
    \omega + E_{nm} + i0} \,,
 }
where
 \eqn{MatrixElements}{
  J_i^{nm} = \langle n | J_i(0, 0) | m \rangle \qquad \vec{P}_{nm} = \vec{P}_n - \vec{P}_m \qquad E_{nm} = E_n - E_m \,,
 }
$\vec{P}_n$ and $E_n$ being the eigenvalues of $\vec{P}$ and $H$ in the state $|n\rangle$.  Plugging \eqref{GijDecomp} into \eqref{sigmaDef} one straightforwardly obtains
 \eqn{sigmaHermitian}{
  {1\over 2} (\sigma_{ij} + \sigma_{ji}^*) = \sum_{n, m} J_i^{nm} J_j^{mn} A_{nm} \qquad
    A_{nm} = {1\over Z} (2 \pi)^3 \delta^{(2)}(\vec{P}_{nm}) \delta (\omega + E_{nm})
    e^{-\beta {E_n + E_m \over 2}} {\sinh {\beta \omega \over 2} \over \omega} \,.
 }
Formally, $A_{nm} \geq 0$, so multiplying \eqref{sigmaHermitian} by an arbitrary column vector $v_j$ to the right and by its adjoint $v_i^*$ to the left yields
 \eqn{vA}{
  v_i^* {1\over 2} (\sigma_{ij} + \sigma_{ji}^*) v_j = \sum_{n, m} \left| v_i^* J_i^{nm} \right|^2 A_{nm} \geq 0 \,,
 }
proving that, indeed, the hermitian part of $\sigma_{ij}$ is positive semi-definite.

The conductivity $\sigma_{ij}(\omega)$ characterizes the response to light of frequency $\omega$ which is incident on the superconductor in a direction normal to the ${\bf R}^{2,1}$ that the sample occupies.  So it is perhaps intuitive that to calculate $\sigma_{ij}(\omega)$ for the black hole, one should send photons down from the boundary of $AdS_4$ and inquire how they are absorbed or reflected by the condensate and the horizon.  More precisely, one uses the gauge-string duality to extract two-point functions from tree-level propagation of photons.  The prescription for computing such Green's functions was first articulated in \cite{Gubser:1998bc,Witten:1998qj}.  An adaptation of it to thermal backgrounds was correctly guessed in \cite{Son:2002sd} and then derived from the original prescription of \cite{Gubser:1998bc,Witten:1998qj} in \cite{Herzog:2002pc} using Schwinger-Keldysh contours.\footnote{Modulo some issues related to behavior near $\omega=0$, this prescription can also be justified \cite{GPR} by the fact that if one analytically continues $G_{ij}^R(\omega)$ to the upper half-plane, then at the Matsubara frequencies $\omega_n = 2 \pi n T$ with $n>0$ it agrees with the corresponding Fourier mode of the Euclidean correlator computed from the prescription proposed in \cite{Gubser:1998bc, Witten:1998qj}.}  In the case of two-point functions, the gauge-string prescription is closely related to D-brane black hole absorption amplitudes computed in a long series of papers beginning with \cite{Das:1996wn}.  If one expresses an asymptotically $AdS_4$ background as
 \eqn{AdSAsymp}{
  ds^2 = {r^2 \over L^2} (-dt^2 + dx^2 + dy^2) + {L^2 \over r^2}
    dr^2 + \hbox{(corrections)} \,,
 }
where the terms shown explicitly are the leading large $r$ behavior, then a complexified photon perturbation polarized in the $x$ direction can be expanded for large $r$ as
 \eqn{PhotonAsymp}{
  A_x = e^{-i\omega t} \left[ A_x^{(0)} + {A_x^{(1)} \over r} +
    O\left( {1 \over r^2} \right) \right] \,,
 }
and the retarded Green's function is given simply by
 \eqn{GRratio}{
  G^R_{xx}(\omega,0) = -{2 \over \kappa^2}
    {A_x^{(1)} \over A_x^{(0)}} \,,
 }
where $\kappa = \sqrt{8\pi G_N}$ is the gravitational coupling, and it is assumed that the photon wave-function is purely infalling at the horizon.  More sophisticated examples have been discussed, for example, in \cite{Hartnoll:2007ai}.

In the superconducting phase of the black holes constructed using the Abelian Higgs model, $\sigma_{xx} = \sigma_{yy}$ and $\sigma_{xy}=0$ because the order parameter is a scalar, breaking gauge invariance but not rotational invariance.  There is a delta-function spike in $\Re\sigma_{xx}(\omega)$ at $\omega=0$, and an associated pole in $\Im\sigma_{xx}(\omega)$.  For non-zero $\omega$ and $T$ not too close to $T_c$, $\Re\sigma_{xx}(\omega)$ is very small up to a finite frequency, which can be denoted $\omega_g = 2\Delta$ in order to evoke a comparison with BCS theory: $\Delta$ is then to be compared with the quantity denoted by the same letter in BCS, whose physical interpretation is the minimal energy of a single normal-component quasi-particle excitation.  Above $\omega_g$, $\Re\sigma_{xx}(\omega)$ rises quickly to a plateau and then asymptotes to a constant as $\omega\to\infty$.  One can argue, along the lines of \cite{Weinberg:1986cq}, that the delta-function spike at $\omega=0$ had to be there because of the broken gauge invariance.  But the existence of a gap is additional information, revealed by the calculations of \cite{Hartnoll:2008vx} but apparently not necessitated by symmetry principles.  In BCS theory, the gap arises because of a pairing mechanism of otherwise nearly free quasi-particle excitations of a Fermi surface.  No such mechanism is manifest in the gravity description; instead, the simplest way to characterize the gravity calculation is that photons with frequency less than $2\Delta$ are very unlikely to penetrate through the condensate and be absorbed by the horizon.  There is clearly something in common between BCS theory and the gravitational calculation, because the horizon represents the dynamics of the uncondensed charge carriers (i.e.~the normal component), and absorption of a photon with $\omega > 0$ is associated with an excitation of these carriers.  The obvious difference is that the charge carriers in the gravitational calculation (or, more precisely, the charge carriers in the appropriate holographic dual description) are strongly coupled even when they are in the normal state.

The strong coupling inherent in gauge-string duals in the gravity approximation raises the appealing possibility that black hole constructions might provide useful physical analogies to the mysterious dynamics of electrons in high $T_c$ materials that go beyond traditional ideas based on quasi-particle excitations of Fermi surfaces.  But the black holes we study provide anything but a microscopic understanding of superconductivity: the gravity description is more like Landau-Ginzburg theory, and the dual field theory would be the venue for some attempt at a microscopic theory comparable to BCS.

Rather than presenting superconducting black holes as an incipient theory of high $T_c$, we prefer the viewpoint that they are a new theoretical laboratory, seemingly divorced from traditional perturbative concepts, but capable of exhibiting assorted phenomena reminiscent of real materials.  Perhaps a sufficiently comprehensive understanding of their dynamics will suggest new ideas which can also be applied successfully to real materials.

The purpose of the present paper is to narrow the gap between black hole constructions and interesting high $T_c$ materials by introducing black holes with a $p$-wave gap.  Although it is apparently a $d$-wave gap that controls the dynamics of the cuprates, $d$-wave and $p$-wave are similar in that excitations of the normal component can be probed using low-frequency photons.

\section{The backgrounds}
\label{BACKGROUND}

We follow the conventions of \cite{Gubser:2008zu} for the metric and gauge field, except for denoting the spatial directions of ${\bf R}^{2,1}$ as $x$ and $y$ rather than $x^1$ and $x^2$.  We will restrict attention to the limit of large $g$, where the metric is simply $AdS_4$-Schwarzschild,
 \eqn{AdSSch}{
  ds^2 = {r^2 \over L^2} \left[ -\left( 1 - {r_H^3 \over r^3}
    \right) dt^2 + dx^2 + dy^2 \right] +
   {L^2 \over r^2} {dr^2 \over 1 - r_H^3/r^3} \,.
 }
The gauge field ansatz is
 \eqn{Abackground}{
  A = \Phi(r) \tau^3 dt + w(r) \tau^1 dx \,.
 }
It is convenient to define
 \eqn{tDefs}{
  \tilde\Phi = gL^2 \Phi \qquad \tilde{w} = gL^2 w \,.
 }
If one also fixes a scale by setting $r_H=1$, then the relevant Yang-Mills equations are
 \eqn{YMbackground}{
  \tilde\Phi'' + {2 \over r} \tilde\Phi' -
    {1 \over r(r^3-1)} \tilde{w}^2 \tilde\Phi &= 0  \cr
  \tilde{w}'' + {1+2r^3 \over r(r^3-1)} \tilde{w}' +
    {r^2 \over (r^3-1)^2} \tilde\Phi^2 \tilde{w} &= 0 \,,
 }
where primes denote $d/dr$.  These equations are similar to (B4) of \cite{Gubser:2008zu} because the ansatz \eno{Abackground} is also similar.  But in \cite{Gubser:2008zu}, the symmetry breaking term takes the form $w (\tau^1 dx + \tau^2 dy)$, which corresponds to wrapping the part of the gauge group generated by $\tau^3$---call it $U(1)_3$---around the rotational symmetry group $SO(2)$ that acts on $x$ and $y$.  Choosing instead $w (\tau^1 dx - \tau^2 dy)$ corresponds to wrapping $U(1)_3$ the other way around $SO(2)$.  We think of \eno{Abackground} heuristically as a superposition of the two different wrapping solutions, in the same way that linearly polarized light is a superposition of left-handed and right-handed polarizations.  This analogy has limited utility because the Yang-Mills equations governing the different ``polarizations'' are non-linear.

In addition to breaking $U(1)_3$, the condensate $w \tau^1 dx$ picks out the $x$ direction as special.  Therefore, if back-reaction of the Yang-Mills field on the metric were included, then we would not expect to be able to set $g_{xx}=g_{yy}$, as we did in \eno{AdSSch}.  The wrapping condensate $w (\tau^1 dx + \tau^2 dy)$ is simpler in this regard, because although it breaks $U(1)_3$ and $SO(2)$ separately, it preserves a diagonal subgroup which makes the stress tensor isotropic in the $x$ and $y$ directions.

The temperature of the horizon is
 \eqn{TDef}{
  T = {3 \over 4\pi L^2} \,,
 }
where, as before, we have set $r_H=1$.  The total charge density $\rho$ is proportional to the $\tau^3$ part of the electric field at infinity: if
 \eqn{PhiExpand}{
  \Phi = p_0 + {p_1 \over r} + O\left( {1 \over r^2} \right) \,,
 }
then
 \eqn{rhoDef}{
  \rho = -{p_1 \over L\kappa^2} \,.
 }
The charge density $\rho_n$ in the normal component is proportional to the $\tau^3$ part of the electric field at the horizon: if
 \eqn{PhiHorizon}{
  \Phi = \Phi_1 (r-1) + O[(r-1)^2] \,,
 }
then
 \eqn{rhonDef}{
  \rho_n = {\Phi_1 \over L \kappa^2} \,.
 }

Far-field and near-horizon expansions for the rescaled fields $\tilde\Phi$ and $\tilde{w}$ take the form
 \eqn{FarT}{
  \hbox{Far field:} &\qquad \left\{ \eqalign{
    \tilde\Phi &= \tilde{p}_0 + {\tilde{p}_1 \over r} + \ldots  \cr
    \tilde{w} &= {\tilde{W}_1 \over r} + \ldots } \right.
 }
 \eqn{NearT}{
  \hbox{Near horizon:} &\qquad \left\{ \eqalign{
    \tilde\Phi &= \tilde\Phi_1 (r-1) + \ldots  \cr
    \tilde{w} &= \tilde{w}_0 + \tilde{w}_2 (r-1)^2 + \ldots \,,}
    \right.
 }
and it is convenient to introduce rescaled versions of the total and normal component charge densities:
 \eqn{RhoTdefs}{
  \tilde\rho &\equiv \kappa^2 gL^2 \rho =
    -{\tilde{p}_1 \over L}  \cr
  \tilde\rho_n &\equiv \kappa^2 gL^2 \rho_n =
    {\tilde\Phi_1 \over L} \,.
 }
We also define the superconducting charge density as $\tilde\rho_s = \tilde\rho - \tilde\rho_n$.  A natural choice of order parameter is $\tilde{W}_1$, because the $SU(2)$ currents $J_m^a$ dual to the gauge bosons $A_\mu^a$ have a symmetry-breaking expectation value
 \eqn{JiaVEV}{
  \langle J_i^a \rangle \propto \tilde{W}_1 \delta^1_i \delta^a_1 \,.
 }
 \begin{figure}
  \centerline{\includegraphics[width=4in]{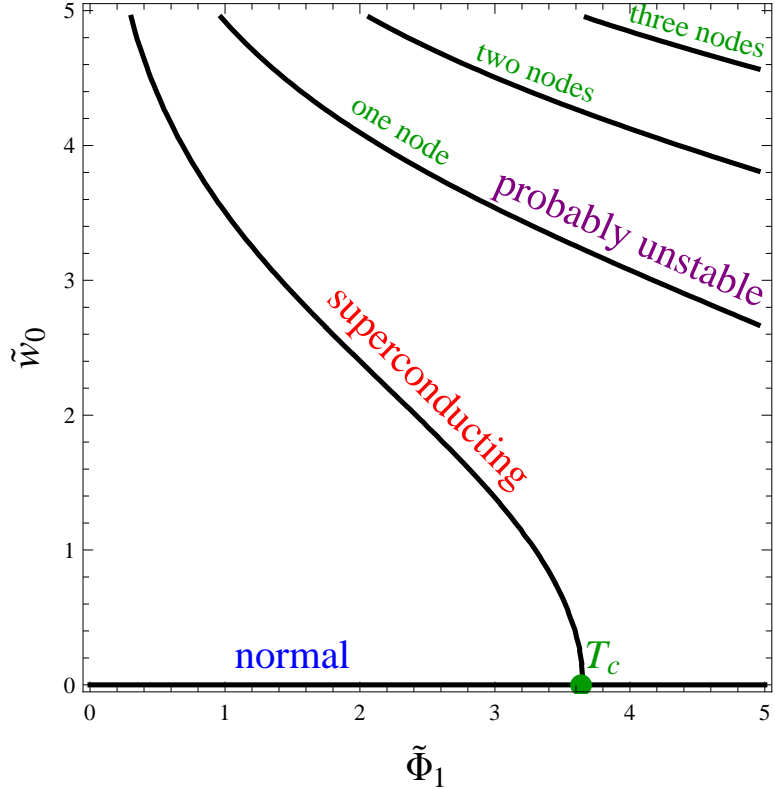}}
  \caption{Each point along the contours plotted represents a solution to the non-linear boundary value problem specified by \eno{YMbackground}, \eno{FarT}, and \eno{NearT}.  Points on the line labeled ``normal'' are RNAdS solutions, and if charge density is held fixed, temperature rises as one moves to the left.  Points on the curve labeled ``superconducting'' break the abelian gauge symmetry generated by $U(1)_3$.  Points on the other curves also break the abelian gauge symmetry but are expected to be unstable.  The point where the superconducting solutions join onto the normal solutions is labeled $T_c$ because the simplest scenario is for there to be a second order phase transition at this point.}\label{SCCONTOURS}
 \end{figure}

There is a one-parameter family of ``allowed'' solutions to the Yang-Mills equations \eno{YMbackground}, where allowed means that the far-field and near-horizon asymptotic forms, \eno{FarT} and \eno{NearT}, are satisfied.  Thus \eno{YMbackground}, \eno{FarT}, and \eno{NearT} specify a non-linear boundary value problem.  To understand why there is only a one-parameter family of solutions, let us examine the far-field and near-horizon expansions separately.  The generic solution to \eno{YMbackground} includes a constant term $\tilde{W}_0$ in the far-field expansion of $\tilde{w}$, and this is disallowed because it corresponds to deforming the field theory lagrangian by some multiple of $J_1^1$.  Another way to describe why $\tilde{W}_0$ is disallowed is that if it is non-zero, then the condensate is not normalizable.  In the expansion \eno{FarT}, all three parameters shown explicitly are independent, which matches a simple counting argument: four integration constants (for two second order differential equations) minus one (for the constraint $\tilde{W}_0=0$) equals three.  Requiring that the gauge field is smooth and well-defined at the horizon leads to the expansions \eno{NearT}.  The parameters $\tilde\Phi_1$ and $\tilde{w}_0$ are independent, but $\tilde{w}_2$ and higher coefficients can be determined in terms of them.  Having only two independent parameters in the near-horizon expansion (i.e.~$\tilde\Phi_1$ and $\tilde{w}_0$) means that there are two constraints at the horizon.  Generically, these two constraints will be independent of the far-field constraint $\tilde{W}_0=0$.  So there are three constraints total on four integration constants, leading indeed to a one-parameter family of solutions.  At special points, one of the horizon constraints may become degenerate with the far-field constraint, and this is when one finds two branches of solutions joining together.
 \begin{figure}
  \centerline{\includegraphics[width=7in]{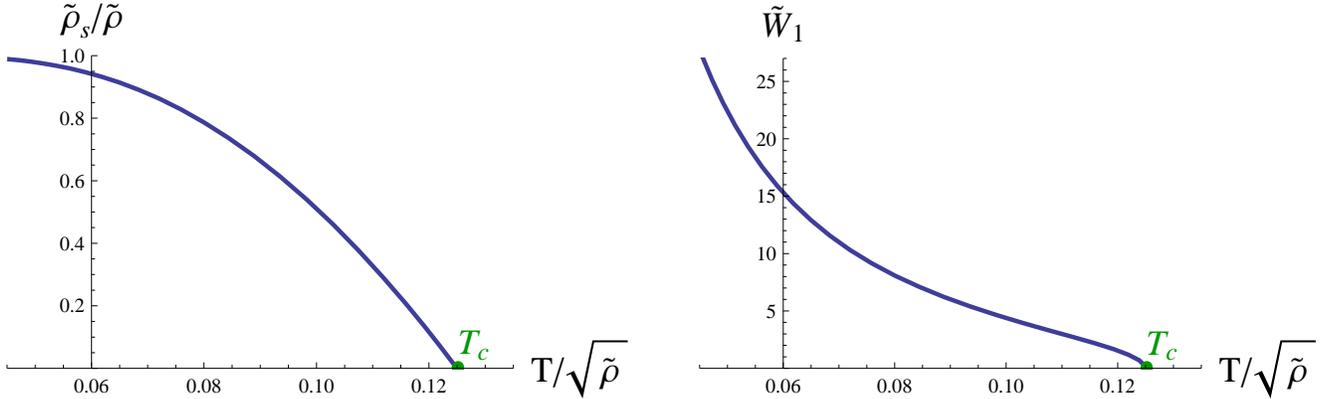}}
  \caption{The fraction $\tilde\rho_s/\tilde\rho$ of the charge carried by the superconducting condensate and the order parameter $\tilde{W}_1$ are plotted against the rescaled temperature $T/\sqrt{\tilde\rho}$.  At $T_c$, $\tilde\rho_s/\tilde\rho$ vanishes linearly, while $\tilde{W}_1$ vanishes as $\sqrt{T_c-T}$.}\label{THERMOPLOTS}
 \end{figure}

Solutions to the boundary value problem discussed in the previous paragraph can be generated using a ``shooting'' procedure.  First one guesses numerical values of $\tilde\Phi_1$ and $\tilde{w}_0$.  Then one uses the near-horizon expansion \eno{NearT} to seed a finite-element differential equation solver, such as Mathematica's {\tt NDSolve}.  Next one matches the numerical solution to the far-field expansion \eno{FarT}, augmented by a constant term $\tilde{W}_0$.  In this way one finds $\tilde{W}_0$ as a function of $\tilde\Phi_1$ and $\tilde{w}_0$.  The zeroes of this function correspond to the solutions of the boundary value problem: see figure~\ref{SCCONTOURS}.  Hereafter we restrict attention to solutions where $\tilde{w}(r)$ has no nodes.  There are additional solutions with nodes, but one generally expects them to be thermodynamically disfavored, because radial oscillations in $\tilde{w}$ cost energy.

Thermodynamic quantities for solutions along the node-free symmetry-breaking branch labeled ``superconducting'' in figure~\ref{SCCONTOURS} are plotted in figure~\ref{THERMOPLOTS}.

\section{Electromagnetic perturbations}
\label{PERTURBATIONS}

By making the black hole charged under the gauge symmetry $U(1)_3$ generated by $\tau^3$, we explicitly break the $SU(2)$ gauge group down to $U(1)_3$.  We interpret $U(1)_3$ as the gauge group of electromagnetism, which means that we plan to consider a weak gauging of this group in the boundary theory.\footnote{This situation is analogous to the Hubbard model, which has (at least) a global $U(2)$ symmetry.  The central $U(1)$ is identified as the electromagnetic gauge symmetry, but electromagnetism is not explicitly part of the model.}  As discussed in the introduction, the linear response to electromagnetic probes is described by the two-point function of the $U(1)_3$ current, and in the dual black hole, this means that we want to know how linear perturbations of the $\tau^3$ component of the gauge field propagate.  We persist in choosing the spatial momenta $k_i=0$ in the $x$ and $y$ directions, so the photon is directed straight down into $AdS_4$, as illustrated in figure~\ref{BALANCE}.

As a warmup, we work out in section~\ref{ANALYTICAL} the conductivity in two examples where it can be done analytically, including the normal state, where the condensate $\tilde{w}$ is set to $0$.  In section~\ref{SETUP} we explain how to set up the perturbation equations in the more difficult case of a symmetry-breaking background as described in section~\ref{BACKGROUND}.  In sections~\ref{NUMERICS} and~\ref{DEPEND} we present results of a numerical study of $\sigma_{xx}(\omega)$ and $\sigma_{yy}(\omega)$ which reveal a $p$-wave gap.

\subsection{Analytical calculations}
\label{ANALYTICAL}

The simplest case to start with is pure $AdS_4$, corresponding to zero temperature, zero charge density, and no symmetry breaking.  At the linearized level, the gauge coupling of $SU(2)$ doesn't enter, so we will pass to free Maxwell theory in $AdS_4$: that is,
 \eqn{FreeMaxwellAdS}{
  S = {1 \over 2\kappa^2} \int d^4 x \, \sqrt{-g}
    \left[ R - {1 \over 4} F_{\mu\nu}^2 + {6 \over L^2} \right] \,.
 }
The perturbation calculation is simple because the background geometry is conformally flat:
 \eqn{JustAdS}{
  ds^2 = {L^2 \over z^2} \left( -dt^2 +
    dx^2 + dy^2 + dz^2 \right) \,.
 }
Equivalently, we may consider the line element \eno{AdSSch} with $r_H=0$: it is the same as \eno{JustAdS} if one sets
 \eqn{rIdent}{
  z=L^2/r \,.
 }
Conformal flatness is special because Maxwell's equations respect conformal symmetry.  Thus the complexified photon perturbation is a plane wave:
 \eqn{PlaneWave}{
  A_x = e^{-i\omega (t-z)} \,.
 }
We chose the plane wave solution that moves in the positive $z$ direction: that is, it moves away from the conformal boundary at $z=0$ and toward the degenerate Killing horizon of the Poincar\'e patch, at $z=\infty$.  (In figure~\ref{BALANCE}, the positive $z$ direction is downward.)  This choice means that we will wind up computing the retarded Green's function rather than the advanced one.  The Green's function can be read off from an expansion near the conformal boundary:
 \eqn{BoundaryExpand}{
  A_x = e^{-i\omega t} (1 + i\omega z + \ldots)
      = e^{-i\omega t} \left( 1 + {i\omega L^2 \over r} + \ldots
        \right) \,.
 }
Comparing the last expression in \eno{BoundaryExpand} to \eno{PhotonAsymp}, and using \eno{sigmaDef} and \eno{GRratio}, one finds
 \eqn{FoundSigmaQ}{
  \sigma_{xx} = \sigma_\infty \equiv {2L^2 \over \kappa^2} \,.
 }
Because of rotation invariance, $\sigma_{yy} = \sigma_{xx}$ and $\sigma_{xy}=0$.  Hereafter we will normalize all conductivities against $\sigma_\infty$ by defining
 \eqn{sigmaTdef}{
  \tilde\sigma_{ij} = {\sigma_{ij} \over \sigma_{\infty}} \,.
 }
Putting \eno{sigmaDef}, \eno{GRratio}, \eno{FoundSigmaQ}, and \eno{sigmaTdef} together, one has
 \eqn{FindSigmaT}{
  \tilde\sigma_{xx} = -{i \over \omega L^2}
   {A_x^{(1)} \over A_x^{(0)}} \,.
 }

A surprising result of \cite{Herzog:2007ij} is that $\tilde\sigma_{xx}=\tilde\sigma_{yy} = 1$ for the $AdS_4$-Schwarzschild solution \eno{AdSSch}, for all $\omega$ and $T$.  In the approximation where the gauge field \eno{Abackground} doesn't back-react on the metric, this result persists so long as $\tilde{w}=0$.  The quickest way to derive it is to compute directly the linearized equation of motions for complexified gauge field perturbations of the background \eno{AdSSch}--\eno{Abackground}: that is, $A \to A + a$, where
 \eqn{EMperturb}{
  a = e^{-i\omega t} a_x^3(r) \tau^3 dx \,.
 }
The result of plugging this perturbation into the linearized Yang-Mills equations is
 \eqn{LinearizedNormal}{
  \left[ \partial_r^2 + {2r^3+1 \over r(r^3-1)} \partial_r +
    {\omega^2 L^4 r^2 \over (r^3-1)^2} \right] a_x^3 = 0 \,,
 }
where we have set $r_H=1$ as usual.  Because the rotational symmetry is unbroken in the absence of the condensate, the same equation governs $a_y^3$ perturbations.  The solution to \eno{LinearizedNormal} describing gauge bosons falling into the horizon at $r=1$ is
 \eqn{NormalAx}{
  a_x^3 = (r-1)^{-i\omega/4\pi T} (r^2+r+1)^{i\omega/8\pi T}
    \left( {r + {1 \over 2} + {i\sqrt{3} \over 2} \over
     r + {1 \over 2} - {i\sqrt{3} \over 2}} \right)^{\sqrt{3}
       \omega / 8\pi T} \,,
 }
where we have used \eno{TDef}.  The behavior $a_x^3 \propto (r-1)^{-i\omega/4\pi T}$ is typical of solutions falling into a finite-temperature horizon.  The expansion of \eno{NormalAx} near the conformal boundary is the same as \eno{BoundaryExpand} through order $1/r$, so the conductivity is the same, as claimed.

\subsection{Electromagnetic perturbations of the superconducting phase}
\label{SETUP}

In the presence of the condensate $\tilde{w} \tau^1 dx$, perturbations of the form \eno{EMperturb} mix with other components at the level of linearized equations.  An ansatz which is sufficiently general to obtain consistent linearized equations is $A \to A + a$, where
 \eqn{Aperturb}{
  a = e^{-i\omega t} \left[
    (a_t^1 \tau^1 + a_t^2 \tau^2) dt +
    a_x^3 \tau^3 dx + a_y^3 \tau^3 dy \right] \,.
 }
All the $a_m^a$ are functions of $r$.  Plugging the perturbation \eno{Aperturb} into the linearized Yang-Mills equations, one finds that the $a_y^3$ mode obeys an equation of motion decoupled from the others:
 \eqn{aySet}{
  \left[ \partial_r^2 + {2r^3+1 \over r(r^3-1)} \partial_r +
    {\omega^2 L^4 r^2 \over (r^3-1)^2} -
     {\tilde{w}^2 \over r(r^3-1)} \right] a_y^3 = 0 \,.
 }
This is identical to (13) of \cite{Hartnoll:2008vx}, except that the last term has slightly different radial dependence.  Unsurprisingly, the rescaled complex conductivity $\tilde\sigma_{yy}$ exhibits similar gapped behavior to what was found in \cite{Hartnoll:2008vx}: see figure~\ref{SIGMAPLOTS}.  Because the analysis is so similar to \cite{Hartnoll:2008vx}, we will not discuss it further here, but simply present the results in sections~\ref{NUMERICS} and~\ref{DEPEND}.

The linearized Yang-Mills equations mix $a_x^3$ with $a_t^1$ and $a_t^2$, resulting in three second order equations of motion,
 \eqn{axEoms}{
  \left[ \partial_r^2 + {2r^3+1 \over r(r^3-1)} \partial_r +
    {\omega^2 L^4 r^2 \over (r^3-1)^2} \right] a_x^3 -
      {r^2 \tilde\Phi \tilde{w} \over (r^3-1)^2} a_t^1 -
      {i\omega L^2 r^2 \over (r^3-1)^2} a_t^2 &= 0  \cr
  \left[ \partial_r^2 + {2 \over r} \partial_r \right] a_t^1 +
    {\tilde\Phi \tilde{w} \over r(r^3-1)} a_x^3 &= 0  \cr
  \left[ \partial_r^2 + {2 \over r} \partial_r -
    {\tilde{w}^2 \over r(r^3-1)} \right] a_t^2 -
    {i\omega L^2 \tilde{w} \over r(r^3-1)} a_x^3 &= 0 \,,
 }
and two first-order constraints,
 \eqn{axConstraints}{
  i\omega L^2 (a_t^1)' + \tilde\Phi (a_t^2)' -
    \tilde\Phi' a_t^2 &= 0  \cr
  i\omega L^2 (a_t^2)' - \tilde\Phi (a_t^1)' +
    \tilde\Phi' a_t^1 -
    \left( 1 - {1 \over r^3} \right)
    \left[ \tilde{w} \partial_r - \tilde{w}' \right] a_x^3 &= 0 \,,
 }
where, as before, primes mean $d/dr$.  The constraints are not independent of the equations of motion: if one takes the $r$ derivative of each constraint, the resulting second order equation follows algebraically from the equations of motion and the undifferentiated constraints.  It takes six constants of integration to specify a solution to the equations of motion, but two of them are used up in satisfying the constraints, leaving four independent solutions.  Of these, two can be found in closed form and are related to residual gauge invariance, as we will discuss in more detail below.  There is also a solution describing gauge bosons falling into the horizon, and another describing gauge bosons coming out.

Let's focus on the infalling solution, which determines a retarded Green's function, as we have seen in easier examples above.  Near the horizon,
 \eqn{NHseries}{
  a_x^3 &= (r-1)^{-i\omega/4\pi T} \left[ 1 +
    a_x^{3(1)} (r-1) + a_x^{3(2)} + \ldots \right]  \cr
  a_t^1 &= (r-1)^{-i\omega/4\pi T} \left[
    a_t^{1(2)} (r-1)^2 + a_t^{1(3)} (r-1)^3 + \ldots \right]  \cr
  a_t^2 &= (r-1)^{-i\omega/4\pi T} \left[
    a_t^{2(1)} (r-1) + a_t^{2(2)} (r-1)^2 + \ldots \right] \,,
 }
and all the coefficients $a_m^{a(s)}$ can be determined once the background and $\omega$ are specified.  Near the conformal boundary, a generic solution to the equations of motion takes the form
 \eqn{FarSeries}{
  a_x^3 &= A_x^{3(0)} + {A_x^{3(1)} \over r} + \ldots \cr
  a_t^1 &= A_t^{1(0)} + {A_t^{1(1)} \over r} + \ldots \cr
  a_t^2 &= A_t^{2(0)} + {A_t^{2(1)} \over r} + \ldots \,,
 }
and the constraints impose the relations
 \eqn{FarConstraints}{
  i\omega L^2 A_t^{1(1)} + \tilde{p}_0 A_t^{2(1)} -
    \tilde{p}_1 A_t^{2(0)} &= 0  \cr
  i\omega L^2 A_t^{2(1)} - \tilde{p}_0 A_t^{1(1)} +
    \tilde{p}_1 A_t^{1(0)} + \tilde{W}_1 A_x^{3(0)} &= 0 \,,
 }
where the coefficients $\tilde{p}_s$ and $\tilde{W}_1$ are the ones appearing in \eno{FarT}.  The infalling solution is unique up to an overall scaling, which is fixed once we choose the leading behavior of $a_x^3$ to be $(r-1)^{-i\omega/4\pi T}$ as in the first line of \eno{NHseries}.  Thus the far-field coefficients $A_m^{a(s)}$ are in principle known as functions of $\omega$ once the background is specified.  We claim that
 \eqn{sigmaXXclaim}{
  \tilde\sigma_{xx} = -{i \over \omega L^2 A_x^{3(0)}} \left(
    A_x^{3(1)} + \tilde{W}_1 {i \omega L^2 A_t^{2(0)} +
    \tilde{p}_0 A_t^{1(0)} \over \tilde{p}_0^2 - \omega^2 L^4}
    \right) \,.
 }
The first term in parentheses is the expected result based on the considerations of \eno{sigmaDef}--\eno{GRratio}.  The second term has to do with solutions to \eno{axEoms} and \eno{axConstraints} which are pure gauge outside the horizon, as we will now explain.

An infinitesimal gauge transformation of the $SU(2)$ gauge field takes the form $\delta A = D\alpha$, where $D = d + gA$ is the gauge-covariant derivative and $\alpha$ is an adjoint scalar gauge function.  Let's consider the case
 \eqn{alphaChoice}{
  \alpha = e^{-i\omega t} \alpha^a \tau^a \,.
 }
After performing the split $A \to A + a$ of the gauge field into background and fluctuating parts, we can view the infinitesimal gauge transformation as acting only on $a = e^{-i\omega t} a_\mu^a \tau^a dx^\mu$:
 \eqn{SmallGauge}{
  \delta (e^{-i\omega t} a_\mu^a) =
    \partial_\mu (e^{-i\omega t} \alpha^a) +
    g \epsilon^{abc} A_\mu^b e^{-i\omega t} \alpha^c \,.
 }
If any $\alpha^a$ depends on $r$, then the gauge-transformed perturbations will include components $a_r^a$ which weren't present in the original ansatz \eno{Aperturb}.  Setting these components to zero amounts to choosing a form of axial gauge, and the gauge transformations that preserve axial gauge are the ones where $\alpha^a$ doesn't depend on $r$.  Dependence on $x^1$ and $x^2$ is excluded because we are always considering modes with zero spatial momentum.  We also set $\alpha^3=0$ because it would introduce components of the perturbations that are not present in the ansatz \eno{Aperturb}.  To summarize: we are interested in infinitesimal gauge transformations of the form \eno{SmallGauge} where $\alpha^1$ and $\alpha^2$ are constant and $\alpha^3=0$.  The explicit form of this restricted set of gauge transformations is
 \eqn{ExplicitGauge}{
  \delta a_x^3 &= \tilde{w} \tilde\alpha^2  \cr
  \delta a_t^1 &= -i \omega L^2 \tilde\alpha^1 - \tilde\Phi
    \tilde\alpha^2  \cr
  \delta a_t^2 &= -i \omega L^2 \tilde\alpha^2 + \tilde\Phi
    \tilde\alpha^1 \,,
 }
where in order to simplify notation we have defined $\tilde\alpha^a = \alpha^a/L^2$.  It is readily checked that the expressions in \eno{ExplicitGauge} solve the equations of motion \eno{axEoms} and the constraints \eno{axConstraints}.  This had to happen because \eno{axEoms}--\eno{axConstraints} came from the gauge-invariant Yang-Mills equations.  These are the two closed-form solutions which we mentioned in the text following \eno{axConstraints}.

Up to an overall scaling, there is a unique linear combination of $a_x^3$, $a_t^1$, and $a_t^2$ which is invariant under the gauge transformation \eno{ExplicitGauge}:
 \eqn{ahatDef}{
  \hat{a}_x^3 = a_x^3 + \tilde{w}
    {i\omega L^2 a_t^2 - \tilde\Phi a_t^1 \over
      \tilde\Phi^2 - \omega^2 L^4} \,.
 }
The conductivity $\tilde\sigma_{xx}$ captures some gauge-invariant information about the bulk theory, and as such it must be expressible in terms of $\hat{a}_x^3$.  If one expands
 \eqn{ahatExpand}{
  \hat{a}_x^3 = \hat{A}_x^{3(0)} + {\hat{A}_x^{3(1)} \over r} +
    \ldots
 }
near the conformal boundary, then the unique extension of \eno{FindSigmaT} that respects the gauge invariance is
 \eqn{ahatConduct}{
  \tilde\sigma_{xx} = -{i \over \omega L^2} {\hat{A}_x^{3(1)} \over
    \hat{A}_x^{3(0)}} \,.
 }
This is precisely the result \eno{sigmaXXclaim} that we claimed earlier.  In appendix~\ref{MATRIX} we describe how $\tilde\sigma_{xx}$ fits into a $3\times 3$ matrix of conductivities which can all be determined in terms of $A^{3(0)}_x$ and $A^{3(1)}_x$.

\subsection{Results of numerics}
\label{NUMERICS}

Let us review the structure of the problem before discussing results.  The gauge field background \eqref{Abackground} is constructed by numerically solving the Yang-Mills equations \eno{YMbackground} in a fixed gravitational background, \eno{AdSSch}, subject to constraints near the conformal boundary and near the horizon, \eno{NearT} and \eno{FarT} respectively.  From a numerical solution, one can pick out the parameters $\tilde{p}_0$, $\tilde{p}_1$, and $\tilde{W}_1$ appearing in \eno{sigmaXXclaim}.  A symmetry-breaking background with $\tilde{w}>0$ everywhere is labeled uniquely by the value of $T/\sqrt{\tilde\rho}$, which has a maximum value of approximately $0.125$.  It is interesting that this value is within numerical error of $1/8$, but we don't see any reason why it should be exactly $1/8$.  With a numerically constructed background in hand, one chooses a value of $\omega$, initializes a finite-element differential equation solver close to the horizon using the series expansion \eno{NHseries}, solves \eno{axEoms}, and extracts the coefficients $A_x^{3(0)}$, $A_x^{3(1)}$, $A_t^{1(0)}$, and $A_t^{2(0)}$ appearing in \eno{sigmaXXclaim} by comparing the far-field behavior of the numerical solution with the expansions \eno{FarSeries}.  It is important to note that $\omega$ and $L$ appear in the differential equations \eno{axEoms} and the conductivity formula \eno{sigmaXXclaim} only in the combination
 \eqn{omegaL}{
  \omega L^2 = {3 \over 4\pi} {\omega \over T} \,,
 }
where we have used \eno{TDef}.  (Recall that we have set $r_H=1$.  If we had not, the left hand side of \eno{omegaL} would be instead $\omega L^2/r_H$, because then $T = 3 r_H / 4\pi L^2$.)  Thus it is more precise to say that one chooses a numerical value for the dimensionless quantity $\omega/T$ and determines $\tilde\sigma_{xx}$, which is also dimensionless, in terms of it.  One expects that for large enough $\omega/T$, $\tilde\sigma_{xx} \to 1$.  This is because the condensate involves dynamics with a characteristic energy scale, which turns out to be $\sqrt{\tilde\rho}$.  Provided we avoid the extreme limit $T \to 0$, $\sqrt{\tilde\rho}$ and $T$ are comparable.  If $\omega \gg \sqrt{\tilde\rho}$, the propagation of the gauge bosons should be largely insensitive to the condensate: instead, its wave function approximately takes the form \eno{PlaneWave} that we found for photons in pure $AdS_4$, and $\sigma_{xx} \approx \sigma_\infty$.

Numerical computations can only detect the continuous part of $\tilde\sigma_{xx}(\omega)$, but there is also a distributional part with some interesting structure.  Because $\tilde\sigma_{xx}(\omega)$ is proportional to a retarded Green's function, it is analytic on the upper half-plane of complex $\omega$.  It therefore satisfies the Kramers-Kronig relations:
 \eqn{KramersKronig}{
  \Re[\tilde\sigma_{xx}(\omega)-1] &=
    {1 \over \pi} {\cal P} \int_{-\infty}^\infty d\omega' \,
    {\Im\tilde\sigma_{xx}(\omega') \over \omega'-\omega}  \cr
  \Im\tilde\sigma_{xx}(\omega) &=
    -{1 \over \pi} {\cal P} \int_{-\infty}^\infty d\omega' \,
    {\Re[\tilde\sigma_{xx}(\omega')-1] \over \omega'-\omega} \,.
 }
The reason that $\tilde\sigma_{xx}-1$ appears in \eno{KramersKronig} rather than $\tilde\sigma_{xx}$ itself is that it is $\tilde\sigma_{xx}-1$ which vanishes as $\omega \to \infty$, and such vanishing is a necessary condition in order to obtain \eno{KramersKronig} from a contour integral in the upper half-plane.  ${\cal P}$ denotes the principle part of the integral.  Evidently, a simple pole in $\Im\tilde\sigma_{xx}(\omega)$ at $\omega=\omega_0$ implies a delta-function contribution $\delta(\omega-\omega_0)$ to $\Re\tilde\sigma_{xx}(\omega)$.  The positivity constraint on the real part of conductivities applies separately to the continuous and delta-function parts of $\Re\tilde\sigma_{xx}(\omega)$, so any pole of $\Im\tilde\sigma_{xx}(\omega)$ on the real axis must have positive residue.

\begin{figure}
   \centerline{\includegraphics[width=7in]{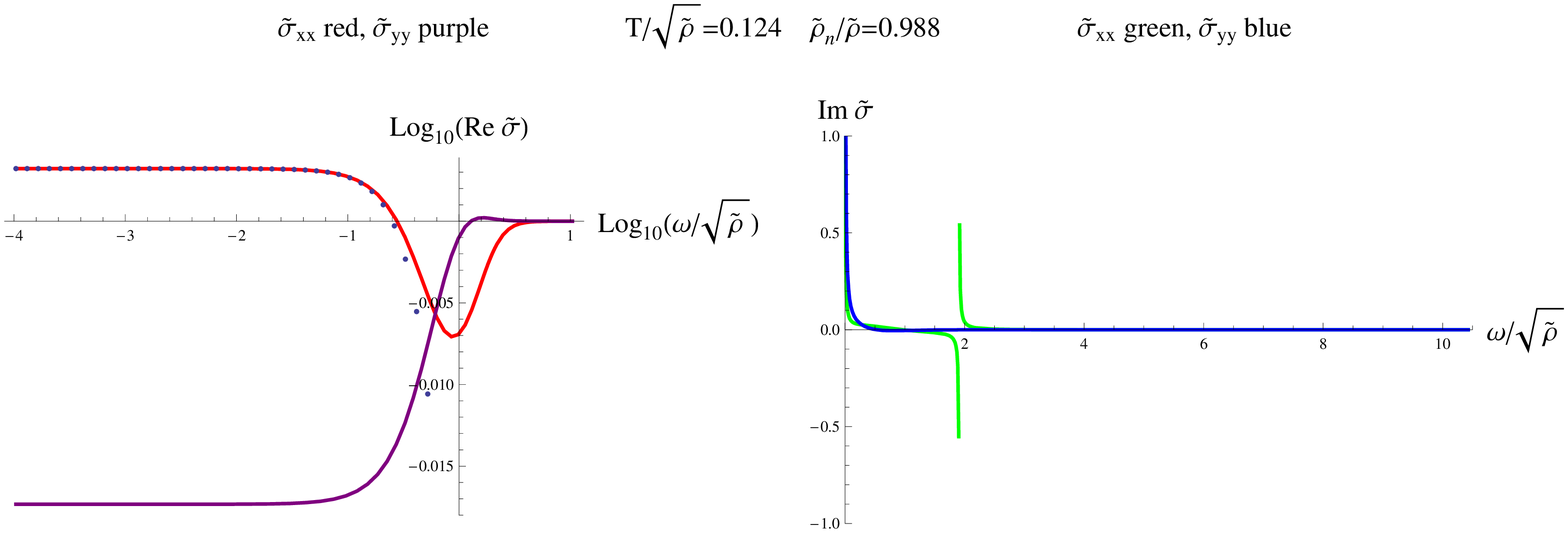}}
   \centerline{\includegraphics[width=7in]{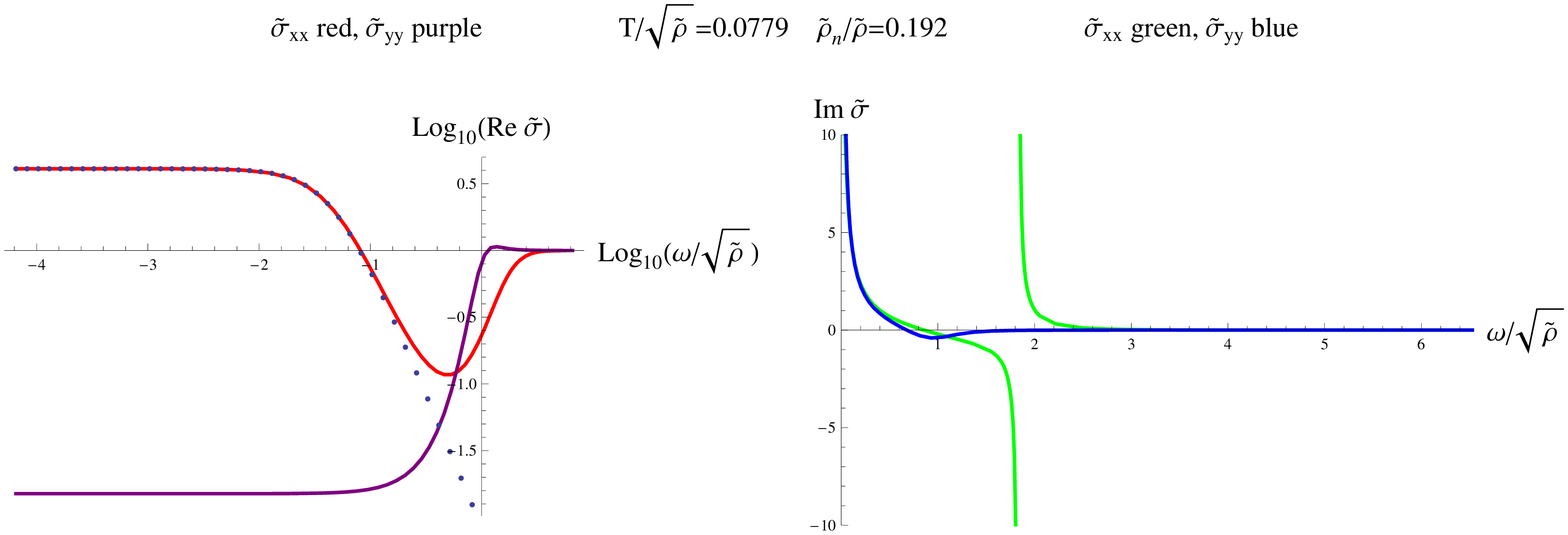}}
   \centerline{\includegraphics[width=7in]{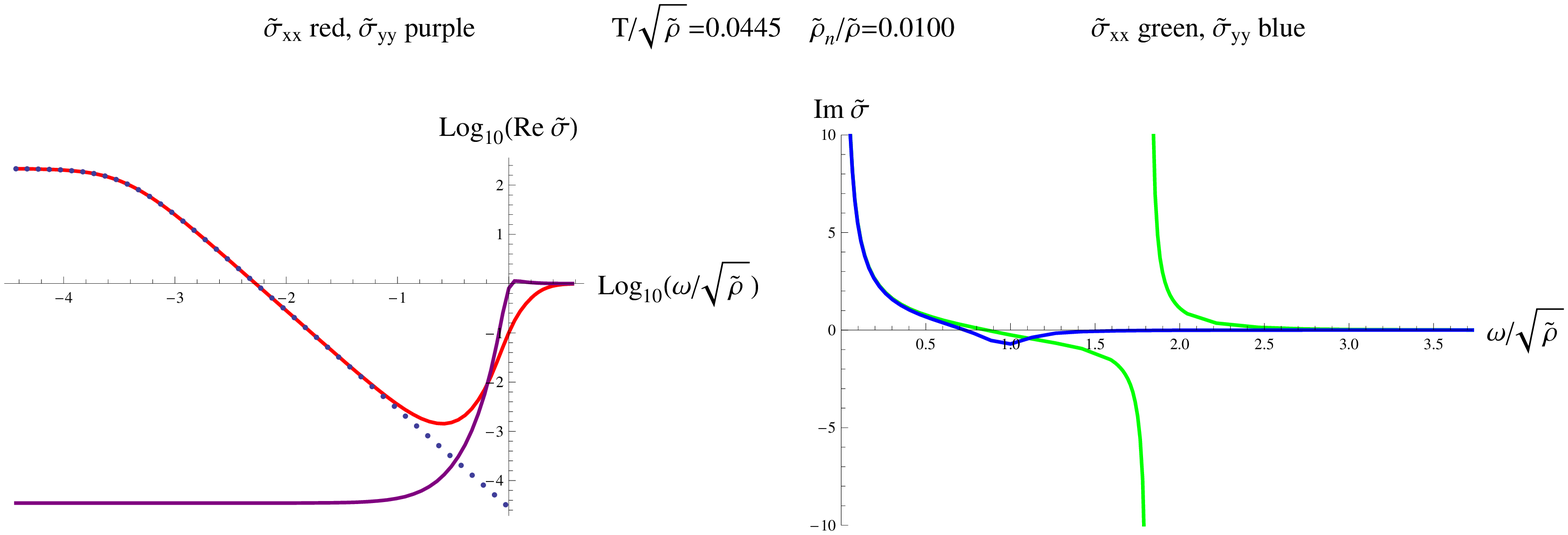}}
  \caption{Rescaled conductivities $\tilde\sigma_{xx}$ and $\tilde\sigma_{yy}$ as functions of frequency.  The dotted curves are the best fits of the Drude model prediction \eno{SigmaDrude} to $\Re\tilde\sigma_{xx}(\omega)$ at low $\omega$.}\label{SIGMAPLOTS}
 \end{figure}

Plots of $\tilde\sigma_{xx}(\omega)$ and $\tilde\sigma_{yy}(\omega)$ are shown in figure~\ref{SIGMAPLOTS}.
The conspicuous features are:
 \begin{enumerate}
  \item $\tilde\sigma_{xx}$ and $\tilde\sigma_{yy}$ both approach $1$ as $\omega$ becomes large, as expected on general grounds.
  \item $\tilde\sigma_{yy}$ displays gapped dependence similar to the findings of \cite{Hartnoll:2008vx}, with $\Delta \approx {1 \over 2} \sqrt{\tilde\rho}$.  That is, $\Re\sigma$ is very small in the infrared, then rises quickly at $\omega = 2\Delta \equiv \omega_g \approx \sqrt{\tilde\rho}$, with a slight ``bump'' a little above $\omega_g$ that is reminiscent of the behavior expected for fermionic pairing.  We use the notation $\omega_g$ even though it's not clear that $\Re\tilde\sigma_{yy}$ is strictly zero for $0 < \omega < \omega_g$.
  \item There is a pole in $\Im\tilde\sigma_{xx}$ at $\omega = \omega_0 \approx 1.8 \sqrt{\tilde\rho}$.  Its residue becomes small as one approaches $T_c$.  It's clear from \eno{sigmaXXclaim} that this pole had to arise, with residue proportional to the order parameter $\tilde{W}_1$: it comes from the denominator of the second term, and
 \eqn{OmegaZero}{
  \omega_0 = {4\pi \over 3} \tilde{p}_0 T \,.
 }
As discussed following \eno{KramersKronig}, there is a delta-function contribution to $\Re\tilde\sigma_{xx}$ at $\omega=\omega_0$, whose coefficient is proportional to the residue of this pole.  This resonance is perfectly stable even at finite temperature, but perhaps if we relax some of the limits we have taken (like large $N$) it would acquire a width.
  \item $\Re\tilde\sigma_{xx}$ never goes as low as $\Re\tilde\sigma_{yy}$, and its rise toward $1$ happens more gradually and at a somewhat larger value of $\omega$, on order $\omega_0$.
  \item The small $\omega$ behavior of $\Re\tilde\sigma_{xx}$ can be parameterized very accurately in terms of the Drude model, which predicts
 \eqn{SigmaDrude}{
  \Re \sigma_{\rm Drude} = {\sigma_0 \over 1 + \omega^2 \tau^2} \,,
 }
where $\sigma_0=ne^2\tau/m$ is a constant related to the density of charge carriers, and $\tau$ is the scattering time.
 \end{enumerate}
We are especially interested in the low-frequency dependence of the conductivities.  Our numerical results make it plausible but not certain that $\sigma_{yy}$ is strictly zero below a finite value of $\omega$ when $T=0$.  However, neglecting the back-reaction of the gauge field may not be a valid approximation for very low temperatures.  On the other hand, the narrow Drude peak in $\tilde\sigma_{xx}$ suggests conductivity due to quasi-particles whose scattering time diverges as $T \to 0$.  Putting the behavior of $\tilde\sigma_{xx}$ and $\tilde\sigma_{yy}$ together suggests a very special type of ``node in the gap,'' namely one which is infinitely narrow as a function of angle in Fourier space.\footnote{We thank D.~Huse and P.~Ong for discussions that led to the picture of an infinitely narrow node described here.}

\subsection{Fits of temperature-dependent quantities}
\label{DEPEND}

In order to extract some simple quantitative information from our numerical results, we considered the dependence of various dimensionless quantities on the rescaled temperature $T/\sqrt{\tilde\rho}$.  Our findings can be summarized briefly as follows:
 \eqn[c]{SeveralFits}{
  {\tilde\rho \over \tilde\rho_n} \approx
    \exp\left\{ 0.303 {\sqrt{\tilde\rho} \over T} - 2.20 \right\}  \cr
  {\tilde{W}_1 \over \tilde\rho} \approx
    1 - 167 \left( {T \over \sqrt{\tilde\rho}} \right)^{3.05}  \cr
  {\tilde\rho_n^2 \over \tilde\rho T} \tau \approx 4.5  \cr
  {\tilde\rho \over \tilde\rho_n^2 \tau^2}
    \lim_{\omega\to 0} \Re\tilde\sigma_{xx}(\omega) \approx 0.302  \cr
  \left( {\tilde\rho \over \tilde\rho_n} \right)^2
    \lim_{\omega\to 0} \Re\tilde\sigma_{yy}(\omega) \approx 0.34  \cr
  \lim_{\omega\to 0} {\omega \over \sqrt{\tilde\rho}}
    \Im\tilde\sigma_{xx}(\omega) \approx 0.52  \cr
  \lim_{\omega\to 0} {\omega \over \sqrt{\tilde\rho}}
    \Im\tilde\sigma_{yy}(\omega) \approx 0.55 \cr
  \lim_{\omega \to \omega_0} {\omega - \omega_0 \over \sqrt{\tilde\rho}}
    \Im\tilde\sigma_{xx}(\omega) \approx 0.28  \,.
 }
The approximately equalities in \eno{SeveralFits} are in some cases quite close over a substantial range of $\sqrt{\tilde\rho}/T$, and in others represent no more than a $T \to 0$ extrapolation: see figure~\ref{Tfits}.  None of the relations \eno{SeveralFits} should be taken too seriously, because they were made over intervals where $T/\sqrt{\tilde\rho}$ varied only by a factor of $5$.
 \def\Res{\mathop{\rm Res}}
 \begin{figure}
  \centerline{\includegraphics[width=5in]{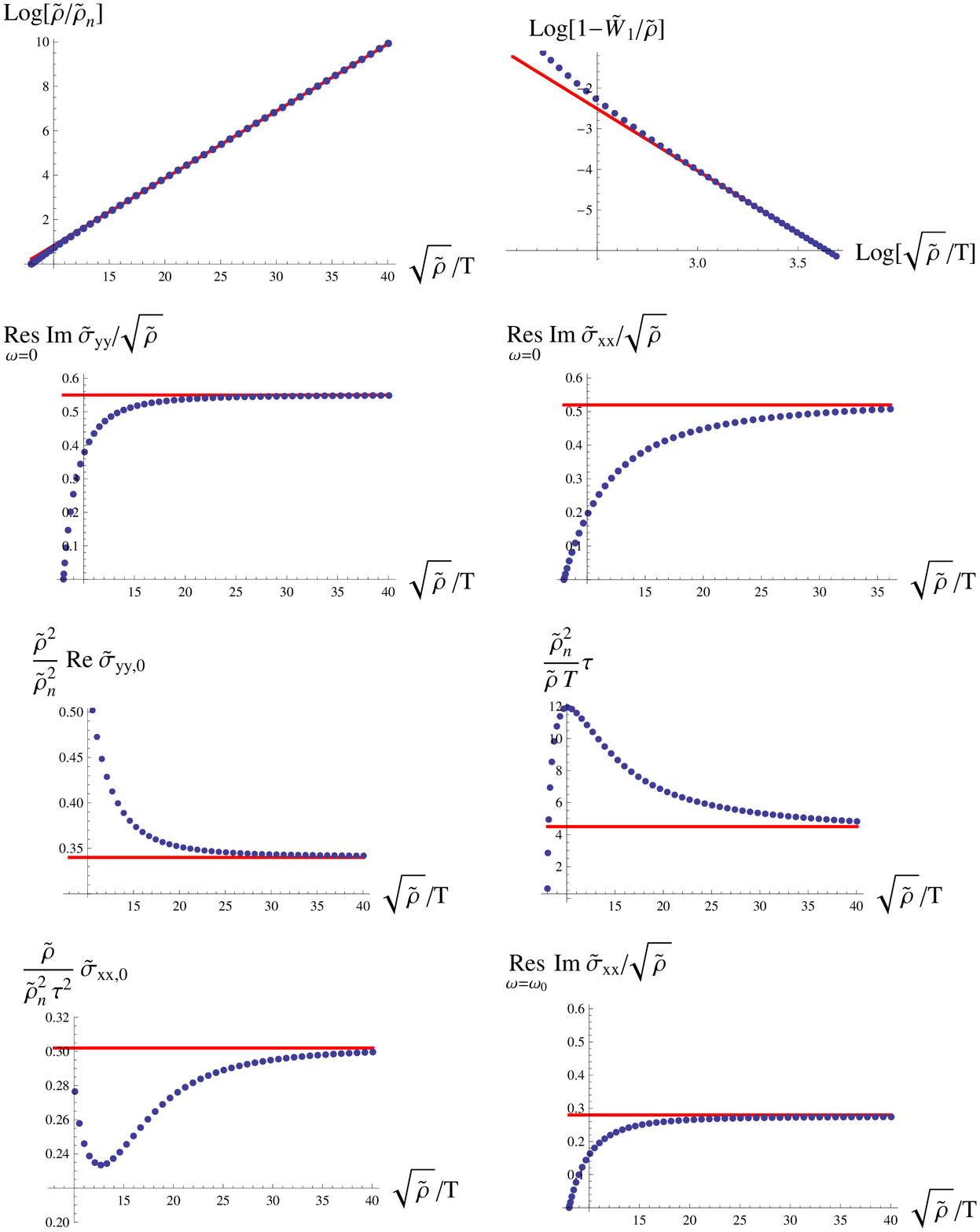}}
  \caption{Temperature-dependent quantities and approximate fits, as explained in \eno{SeveralFits} and the surrounding text.  We have defined $\tilde\sigma_{xx,0} = \lim_{\omega\to 0} \Re\tilde\sigma_{xx}(\omega)$, $\tilde\sigma_{yy,0} = \lim_{\omega\to 0} \Re\tilde\sigma_{yy}(\omega)$, $\Res_{\omega = 0} \Im\tilde\sigma_{xx}/\sqrt{\tilde\rho} = \lim_{\omega \to 0} {\omega \over \sqrt{\tilde\rho}} \Im\tilde\sigma_{xx}(\omega)$, $\Res_{\omega = 0} \Im\tilde\sigma_{yy}/\sqrt{\tilde\rho} = \lim_{\omega \to 0} {\omega \over \sqrt{\tilde\rho}} \Im\tilde\sigma_{yy}(\omega)$, and $\Res_{\omega = \omega_0} \Im\tilde\sigma_{xx}/\sqrt{\tilde\rho} = \lim_{\omega \to \omega_0} {\omega-\omega_0 \over \sqrt{\tilde\rho}} \Im\tilde\sigma_{xx}(\omega)$.}\label{Tfits}
 \end{figure}
A particularly challenging case is the quantity $\lim_{\omega \to 0}{\omega \over \sqrt{\tilde\rho}} \Im\tilde\sigma_{xx}(\omega)$.  The $\omega \to 0$ limit converges slowly because of a ``shelf effect:'' for values in a region around $\omega \sim 1/\tau$, we observed ${\omega \over \sqrt{\tilde\rho}} \Im\tilde\sigma_{xx} \approx 0.55$ at low temperatures, which is the same value as we find in the $\omega \to 0$ limit for ${\omega \over \sqrt{\tilde\rho}} \Im\tilde\sigma_{yy}$.  But for $\omega \lsim 1/50\tau$, we observed instead the value some $6\%$ smaller quoted in \eno{SeveralFits}.  Our numerical algorithms aren't optimized for extremely small $T$ and $\omega$, and it's possible that this shelf effect goes away at very small $T$, so that the residues of $\Im\tilde\sigma_{xx}$ and $\Im\tilde\sigma_{yy}$ agree in this limit.  But the balance of evidence from our numerical exploration is that this does not happen, or happens very slowly as $T$ is decreased.

\section{Stability calculations}
\label{STABILITY}

We expected that the $p$-wave backgrounds \eno{Abackground} would be unstable against small perturbations that would eventually turn them into backgrounds of the type studied in \cite{Gubser:2008zu}.  These backgrounds display behavior analogous to a $p+ip$ gap.\footnote{The analogy to a $p+ip$ gap is apt because the combination $\tau^1 dx + \tau^2 dy$ distinguishes an orientation on ${\bf R}^2$ and implies a spontaneous magnetization.  To see this, note first that the positive charge of the black hole under $U(1)_3$ privileges $\tau^3$ over $-\tau^3$.  The structure constants $\epsilon^{abc}$ of $SU(2)$ then privilege the ordering $(\tau^1,\tau^2)$ over $(\tau^2,\tau^1)$, because having distinguished the positive $\tau^3$ direction in the Lie algebra lets us set $c=3$.  Finally, $\tau^1 dx + \tau^2 dy$ ``locks'' this orientation in the Lie algebra to an orientation $dx \wedge dy$ on ${\bf R}^2$.  More physically, a contribution $w (\tau^1 dx + \tau^2 dy)$ to $A$ means that there is a term $w^2 \tau^3 dx \wedge dy$ in $F$, representing a spontaneous magnetization that again picks out an orientation $dx \wedge dy$ in ${\bf R}^2$.  In any case, the symmetries of this state are clearly those of a $p+ip$ gap whose $ip$ component is of identical magnitude to its $p$ component, so that the gap is uniform in magnitude but has a phase that rotates by $2\pi$ as one goes once around the Fermi surface.}  But the opposite seems to be true: numerical explorations of quasinormal modes close to $T_c$ show that it is the $p+ip$-wave backgrounds that are unstable, and it seems that they evolve toward pure $p$-wave backgrounds, which are stable.  In section~\ref{STABLE} we exhibit the equations describing the perturbations of the pure $p$-wave backgrounds that we thought would be unstable and explain how the lowest-lying quasinormal modes exhibit stability instead, close to $T_c$.  In section~\ref{PPLUSIP}, we show that similar perturbations of the backgrounds studied in \cite{Gubser:2008zu} exhibit an instability slightly below $T_c$.

\subsection{Quasinormal frequencies of $p$-wave backgrounds}
\label{STABLE}

Let us begin by explaining why we thought $p$-wave backgrounds would be unstable. At $T=T_c$, both the $\tau^1 dx$ mode and the $\tau^2 dy$ directions exhibit marginally stable modes.  So a natural expectation is that both become unstable for $T<T_c$.  Yet the $p$-wave backgrounds described in section~\ref{BACKGROUND} involve only $\tau^1 dx$, whereas the $p+ip$-wave backgrounds of \cite{Gubser:2008zu} involve the combination $\tau^1 dx + \tau^2 dy$.  In the latter case we are taking advantage of both directions of instability, and it seems reasonable that such a configuration should be preferred.  But this reasoning ignores the non-linearities of the Yang-Mills equations.  It turns out that condensing in the $\tau^1 dx$ direction stabilizes against condensation in the $\tau^2 dy$ direction---at least, close to $T_c$.  That stabilization is what we are going to address in this section.
 \begin{figure}
   \centerline{\includegraphics[width=4.5in]{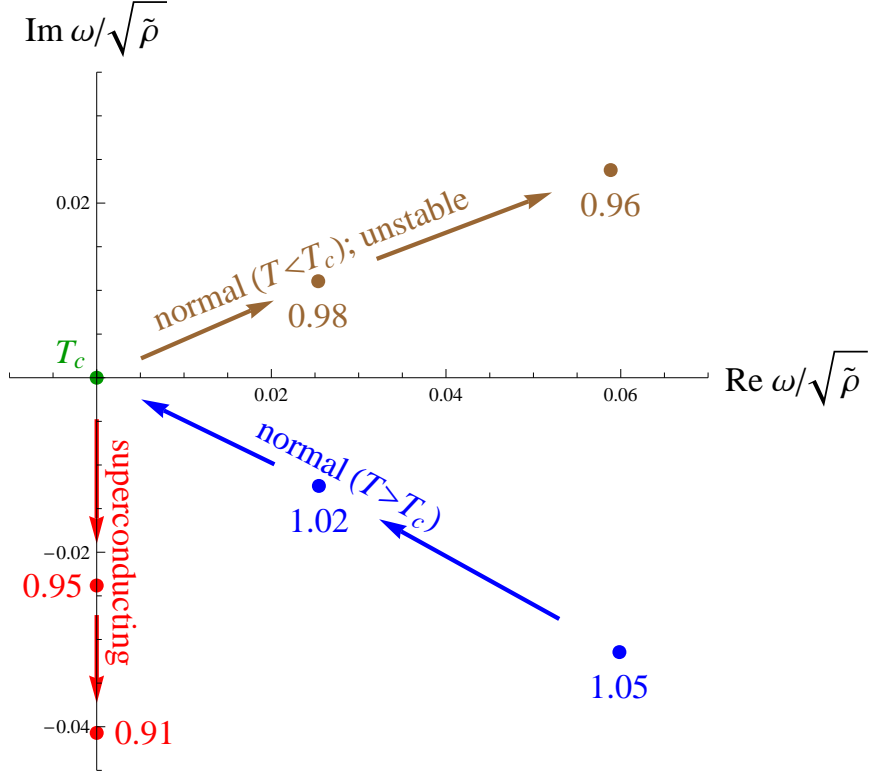}}
  \caption{Quasinormal frequencies corresponding to the perturbation \eqref{ayeoms} of the $p$-wave superconducting background \eqref{Abackground} near the critical temperature.  The quasinormal mode spectrum is symmetric about the imaginary axis, and we are only showing the quasinormal frequencies with non-negative real parts.  The arrows are in the direction of decreasing temperature, and the number displayed next to each quasinormal frequency represents $T/T_c$.  The blue points correspond to backgrounds with no condensate above $T_c$; the brown points correspond to backgrounds with no condensate below $T_c$; and the red points correspond to superconducting backgrounds below $T_c$.  The superconducting backgrounds also have a quasinormal mode at $\omega = 0$ (see main text) which is not displayed.  The backgrounds with no condensate below $T_c$ have quasinormal frequencies with positive imaginary parts, indicating an instability.  The other backgrounds (namely normal state above $T_c$ and superconducting below $T_c$) appear to be stable.}\label{QNMp}
 \end{figure}

Starting from the backgrounds \eno{Abackground}, we want to study $\tau^2 dy$ perturbations, which is to say $a_y^2$.  At the linearized level, $a_y^2$ couples with $a_y^1$, so we are forced to examine the combined perturbation $A \to A + a$, where
 \eqn{ayPert}{
  a = e^{-i\omega t}
    \left( a_y^1 \tau^1 + a_y^2 \tau^2 \right) dy \,.
 }
The equations of motion read
 \eqn{ayeoms}{
  \left[\partial_r^2 + {2r^3 + 1 \over r(r^3 - 1)} \partial_r
    + {r^2(\omega^2 L^4 + \tilde\Phi^2) \over (r^3 - 1)^2} \right] a_y^1
    - {2 i \omega L^2 r^2 \tilde\Phi \over (r^3 - 1)^2} a_y^2 &= 0 \cr
  \left[\partial_r^2 + {2r^3 + 1 \over r(r^3 - 1)} \partial_r
    + {r^2(\omega^2 L^4 + \tilde\Phi^2) \over (r^3 - 1)^2} -
    {\tilde w^2 \over r(r^3 - 1)}\right] a_y^2
    + {2 i \omega L^2 r^2 \tilde\Phi \over (r^3 - 1)^2} a_y^1 &= 0 \,.
 }
The appropriate boundary conditions for quasinormal modes are that $a_y^1$ and $a_y^2$ should vanish at the boundary of $AdS_4$ and that $a$ should be a function only of the infalling coordinate $t+{1 \over 4\pi T} \log(r-1)$ at the black hole horizon (where, as usual, $r_H=1$).  These conditions can be simultaneously satisfied only for certain complex quasinormal frequencies $\omega$.  Since we assumed $e^{-i \omega t}$ time dependence, quasinormal frequencies with negative imaginary parts correspond to stable modes, while those with positive imaginary parts correspond to unstable modes.  Solutions with purely real $\omega$ correspond to true normal modes of the system.  From the symmetries of the equations \eqref{ayeoms} and of the boundary conditions described above, it follows that if $\omega$ is a quasinormal frequency, then so is $-\omega^*$.  So let's restrict attention to quasinormal frequencies with $\Re\omega \geq 0$.  Figure~\ref{QNMp} shows how the lowest-lying quasinormal frequencies behave as functions of temperature close to $T_c$.  Above $T_c$, the normal state is stable, and the quasinormal modes come in degenerate pairs with the same $\omega$.  As we mentioned earlier, there are two quasinormal modes that become marginally stable at $T_c$: their frequencies go to zero.  One of these modes, involving only $a_y^1$, stays right at $\omega=0$ below $T_c$ on the superconducting branch.  It is a Goldstone mode describing spatial rotations of the condensate.  The other mode is stable on the superconducting branch below $T_c$.  What makes it stable is the $-\tilde{w}^2 \over r(r^3-1)$ term in the second equation of \eno{ayeoms}.  This term is like a positive, $r$-dependent contribution to the mass term of the gauge boson.  Dropping this term amounts to passing to the normal state below $T_c$, and our normal investigation showed that this state is unstable.  So the $-\tilde{w}^2 \over r(r^3-1)$ term is the advertised stabilization mechanism, and it is evidently due to the non-linearities of the Yang-Mills equations of motion.

\subsection{Quasinormal frequencies of $p + ip$-wave backgrounds}
\label{PPLUSIP}

We now wish to show that the large $gL$ limit of the $p+ip$ backgrounds studied in \cite{Gubser:2008zu} are unstable, at least for $T$ close to $T_c$.  The instability decreases the $ip$ component of the gap and appears likely to lead the system into a $p$-wave state like \eno{Abackground}.  Our strategy is to find out what happens to the modes which are marginally stable at $T_c$ as we go slightly away from the critical temperature on the superconducting and normal branches.

At large $g$, the gauge field ansatz for the circularly polarized backgrounds is
 \eqn{ApPlusIp}{
  A = \Phi(r) \tau^3 dt + w(r) \left(\tau^1 dx + \tau^2 dy \right) \,,
 }
and it is again convenient to define
 \eqn{tDefsAgain}{
  \tilde\Phi = gL^2 \Phi \qquad \tilde{w} = gL^2 w \,.
 }
In the large $g$ limit there is no back-reaction on the metric, so the metric is simply \eqref{AdSSch}.  The equations of motion for $\tilde \Phi$ and $\tilde w$ are similar to \eqref{YMbackground}.  They are given explicitly in~(B4) of \cite{Gubser:2008zu}, and we will not reproduce them here.
 \begin{figure}
   \centerline{\includegraphics[width=4.5in]{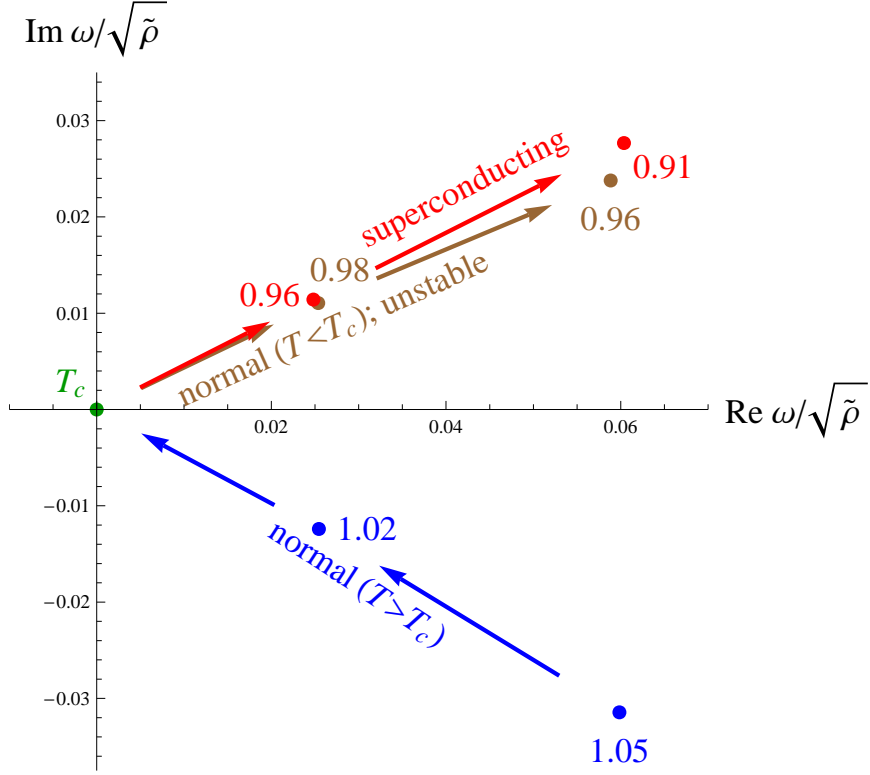}}
  \caption{Quasinormal frequencies corresponding to the perturbation \eqref{aCircPert} of the $p+ip$-wave background \eqref{ApPlusIp} near the critical temperature.  The spectrum of quasinormal modes is symmetric about the imaginary axis, and we are only showing the quasinormal frequencies with non-negative real parts.  The arrows are in the direction of decreasing temperature, and the number displayed next to each quasinormal frequency represents $T/T_c$.  The blue points correspond to backgrounds with no condensate above $T_c$; the brown points correspond to backgrounds with no condensate below $T_c$; and the red points correspond to superconducting backgrounds below $T_c$.  The backgrounds with no condensate above $T_c$, as well as the superconducting ones below $T_c$, have quasinormal frequencies with positive imaginary parts, indicating an instability.  The backgrounds with no condensate above $T_c$ are likely to be stable.}\label{QNMpCLG}
 \end{figure}

There are many ways in which one can perturb the background \eqref{ApPlusIp}, but the perturbations that might show an instability towards converting $p+ip$ into $p$ should be of the form
 \eqn{aCircPert}{
  a = e^{-i \omega t} a_1 (\tau^1 dx - \tau^2 dy) + e^{-i \omega t} a_2 (\tau^2 dx + \tau^1 dy) \,.
 }
The $a_1$ perturbation changes the relative magnitude of the $p$ and $ip$ components of the background ansatz \eno{ApPlusIp}.  Nothing in the ansatz \eno{ApPlusIp} picks out whether $\tau^1 dx$ or $\tau^2 dy$ is the $p$-wave part (as opposed to the $ip$ part) so changing the relative size of these two components with a linear perturbation can be interpreted as decreasing the $ip$ component without loss of generality.  The $a_2$ component is a $90^\circ$ spatial rotation of the $a_1$ component.  The linearized equations for $a_1$ and $a_2$ are
 \eqn{a12eoms}{
  \left[\partial_r^2 + {2r^3 + 1 \over r(r^3 - 1)} \partial_r
    + {r^2(\omega^2 L^4 + \tilde\Phi^2) \over (r^3 - 1)^2}
    + {\tilde w^2 \over r(r^3 - 1)} \right] a_1
    + {2 i \omega L^2 r^2 \tilde\Phi \over (r^3 - 1)^2} a_2 &= 0 \cr
  \left[\partial_r^2 + {2r^3 + 1 \over r(r^3 - 1)} \partial_r
    + {r^2(\omega^2 L^4 + \tilde\Phi^2) \over (r^3 - 1)^2}
    + {\tilde w^2 \over r(r^3 - 1)} \right] a_2
    - {2 i \omega L^2 r^2 \tilde\Phi \over (r^3 - 1)^2} a_1 &= 0 \,.
 }
The perturbations should take the form of infalling waves close to the horizon and should vanish at the boundary, as in the case of the linearly polarized backgrounds examined in the previous section.  Only for discretely many quasinormal frequencies are these boundary conditions satisfied.

When $\tilde w = 0$, equations \eqref{a12eoms} are the same as the equations for $a_y^1$ and $a_y^2$ given in \eqref{ayeoms}, so at zero condensate the quasinormal modes coincide with the ones displayed in figure~\ref{QNMp}.  When $T<T_c$ we find an instability whether or not there is a condensate: see figure~\ref{QNMpCLG}.  This result could perhaps have been anticipated by noting that the $\tilde{w}^2/r(r^3-1)$ terms in \eno{a12eoms} enter with the opposite sign from the way they entered \eno{ayeoms}.  So instead of tending to stabilize perturbations, they tend to destabilize them.  It's worth noting, however, that $\tilde{w}$ is the coefficient of $\tau^1 dx$ in \eno{ayeoms}, whereas it is the coefficient of $\tau^1 dx + \tau^2 dy$ in \eno{a12eoms}.  So, the functional forms of $\tilde{w}$ will differ in the two cases, becoming equal only in the limit $T \to T_c$.

\section{Discussion}
\label{DISCUSSION}

The distinguishing feature of the superconducting black holes constructed in this paper is that the condensate is anisotropic, in the sense of picking out the $x$ direction as preferred.  This is in contrast to earlier constructions \cite{Gubser:2005ih,Gubser:2008px,Hartnoll:2008vx,Gubser:2008zu}.  What is special about the $x$ directions is that the conductivity in this direction, $\sigma_{xx}$, becomes large at small but non-zero $\omega$.  So far, the situation is similar to $p$-wave superconductors.  But in real materials, impurity scattering would keep $\sigma_{xx}$ finite for small non-zero $\omega$, whereas in our setup, the only upper bound comes from the effects of finite temperature.  The biggest difference from real materials---from the perspective of the electromagnetic response---is that $\sigma_{yy}$ displays gapped dependence, similar to what was found for an $s$-wave construction in \cite{Hartnoll:2008vx}.  In real $p$-wave materials, the gap vanishes at $\theta=0$ and $\theta=\pi$ but has finite slope there.  Gapped $\sigma_{yy}$ suggests instead an infinitely narrow node in the gap: the slope of $\Delta$ as a function of $\theta$ is infinite at $\theta=0$ and $\pi$.  To put it another way, the states which usually occupy a Dirac cone near a $p$-wave gap have been squeezed into a purely one-dimensional structure, at least in the limit of low energy.  We emphasize that this picture of an infinitely narrow node in the gap is entirely heuristic, given that we do not have a microscopic description of the condensate in the language of a dual CFT\@.  What we can say most clearly in the CFT language is that there is an $SU(2)$ current algebra, and when there is a strong enough chemical potential for the charge density $J_t^3$, the component $J^1_x$ develops an expectation value.  We are tempted to conjecture that $J^a_m \sim \bar\psi_i \gamma_m \tau^a_{ij} \psi_i$ for some fermion fields $\psi_i$ in a representation of $SU(2)$.  Then the condensate is composed of fermion pairs created by $J^1_x$, which have one unit of angular momentum.

Our results are preliminary in various ways:
 \begin{enumerate}
  \item We didn't consider back-reaction of the gauge field on the metric.  Back-reaction can be suppressed by taking the gauge coupling large, but this limit is non-uniform in that as $T \to 0$, the $A_x^1$ component of the gauge field gets larger and larger at the horizon, demanding a bigger value of the gauge coupling to justify the neglect of back-rection.
  \item Our conductivity calculations do not allow for spatial momentum.  In other words, we calculated a retarded two-point function $G_R(\omega,0)$ of $J^3_i$ at non-zero frequency but zero spatial momentum.  A study of the electromagnetic response at non-zero $k$ might help consolidate the heuristic Fermi-surface picture we have offered, or it might invalidate it and suggest a different interpretation.
  \item We encountered some curious numerical coincidences, ranging from $T_c/\sqrt{\tilde\rho} \approx 1/8$ to the scaling of the ``scattering rate'' $1/\tau$ and the small $\omega$ limits of $1/\tilde\sigma_{xx}$ and $\tilde\sigma_{yy}$ approximately as $\tilde\rho_n^2$ rather than some fractional power of $\tilde\rho_n$.  The latter coincidence evokes the idea that the behavior of quantities like the scattering rate are largely controlled by kinematic factors of two incoming quasi-particles.  It would be interesting if some of these numerical coincidences could be understood in terms of exact solutions to the Yang-Mills equations, or in terms of some systematic approximation scheme rather than brute-force numerics.
  \item The scope of our stability calculations is very restricted: not only have we limited ourselves to the no-back-reaction limit, but we also stayed close to $T_c$.  Moreover, we do not claim to have considered every possible perturbation, only the ones that seemed obvious candidates for exhibiting instabilities.  It would clearly be desirable to be more thorough.
  \item String theory or M-theory compactified on a manifold of positive curvature leads, often if not typically, to a theory of gauged supergravity which would include the Einstein-Yang-Mills lagrangian, with particular relations between the gauge coupling and the cosmological constants enforced by supersymmetry (assuming there is supersymmetry).  Does superconductivity occur in any such construction?  Is it the gauge bosons which condense first, or are there charged scalars which condense?  Is there a higher-dimensional interpretation---for example, some sort of Gregory-Laflamme instability involving the compact dimensions?
  \item The constructions we have discussed probably generalize to higher dimensions.
  \item We have limited ourselves entirely to classical configurations, excluding any discussion of fluctuations.  This would seem problematic in two spatial dimensions because of infrared divergences, but fluctuations are suppressed when the radius of $AdS_4$ is much larger than the Planck scale, corresponding to a large $N$ limit in the dual CFT\@.  But to understand the condensate's contribution to the specific heat, presumably one should consider fluctuations.
 \end{enumerate}
We hope to report on some of these issues in the future.

\section*{Acknowledgments}

We thank A.~Bernevig, W.~Brinkman, S.~Hartnoll, P.~Ong, J.~Preskill, M.~Roberts, and especially D.~Huse for useful discussions.  This work was supported in part by the Department of Energy under Grant No.\ DE-FG02-91ER40671 and by the NSF under award number PHY-0652782.

\clearpage

\appendix

\section{Conductivity and resistivity matrices}
\label{MATRIX}

The conductivities $\sigma_{xx}$ and $\sigma_{yy}$ that we computed in section~\ref{PERTURBATIONS} are the $\sigma_{xx}^{33}$ and $\sigma_{yy}^{33}$ entries of a generalized conductivity matrix $\sigma_{mn}^{ab}$ given by
 \eqn{sigmaDefAgain}{
  \sigma_{mn}^{ab}(\omega) = {i \over \omega} G^{R, ab}_{mn}(\omega,0) \,,
 }
where
 \eqn{GRdefAgain}{
  G^{R, ab}_{mn}(\omega,\vec{k}) = -i \int d^4 x \,
    e^{i\omega t - i \vec{k} \cdot \vec{x}}
    \theta(t) \langle [J_m^a(t,\vec{x}),J_n^b(0,0)] \rangle \,.
 }
The matrix $\sigma_{mn}^{ab}$ is block-diagonal, each block corresponding to a set of gauge perturbations in the bulk that mix with each other.  To compute $G^{R, ab}_{mn}$ from the bulk, it is useful to pass to a complexified gauge perturbation, $A \to A + a$, and write the field strength perturbation as $F \to F + f$.  Then $f = Da$, where $D=d+gA$ is the gauge-covariant derivative, and we need to work only to leading order in $a$.  The quadratic action is
 \eqn{QuadraticAction}{
  S_2 = {1\over 2 \kappa^2} \int d^4 x \, \sqrt{-g} \left( -{1 \over 2}
    f_{\mu\nu}^{a*} f^{\mu\nu a} \right) \,.
 }
From now on let's focus on the $3 \times 3$ block corresponding to $J_t^1$, $J_t^2$, and $J_x^3$---in other words, we keep only $a_t^1$, $a_t^2$, and $a_x^3$ non-zero and set all the other $a_m^a$ to zero.  Integrating \eqref{QuadraticAction} by parts, we obtain
 \eqn{OnshellAction}{
  S_2 = {1\over 2 \kappa^2} \int d^4 x \, \left[ a_\mu^{a*} \left( \hbox{eom for $a_\mu^a$}
    \right) + \partial_r {\cal J} \right] \,,
 }
where
 \eqn{JDef}{
  {\cal J} = {1\over 2 \kappa^2} {r^2 \over L^2} \left( a_t^{1*} \partial_r a_t^1 +
    a_t^{2*} \partial_r a_t^2 \right) +
   {1\over 2 \kappa^2} {1-r^3 \over L^2 r} a_x^{3*} \partial_r a_x^3 \,.
 }
We henceforth set $L=1$ to further simplify notation.

As explained in section~\ref{SETUP}, we can find a solution to the equations of motion and two constraints for $a_t^1$, $a_t^2$, and $a_x^3$ (that is, equations \eno{axEoms} and \eno{axConstraints}) which is unique up to an overall multiplicative factor.  The uniqueness arises because there's only one infalling solution at the horizon.  But this purely infalling solution may not have the desired properties at the boundary of AdS, so we should consider gauge-equivalents of it:
 \eqn{GaugeEquivalents}{
  a_t^1 &\to \tilde\alpha^0 a_t^1 -
    i\omega \tilde\alpha^1 - \tilde\Phi
    \tilde\alpha^2  \cr
  a_t^2 &\to \tilde\alpha^0 a_t^2 -
    i\omega \tilde\alpha^2 + \tilde\Phi
    \tilde\alpha^1  \cr
  a_x^3 &\to \tilde\alpha^0 a_x^3 + \tilde{w} \tilde\alpha^2 \,.
 }
The ``gauge parameter'' $\tilde\alpha^0$ is just a multiplicative rescaling factor, which we are free to include.  $\tilde\alpha^1$ and $\tilde\alpha^2$ should be independent of $r$ at least on some neighborhood of the boundary.

In order to compute two-point functions of the $J_i$'s, we want to prescribe the value of the gauge-transformed perturbations at the boundary by sending
 \eqn{ChooseatOneBdy}{
  \tilde\alpha^0 a_t^1 - i\omega L^2 \tilde\alpha^1 -
    \tilde\Phi \tilde\alpha^2 \quad &\to\quad \beta_t^1
    \equiv -\beta^t_1  \cr
  \tilde\alpha^0 a_t^2 - i\omega L^2 \tilde\alpha^2 +
    \tilde\Phi \tilde\alpha^1 \quad &\to\quad \beta_t^2
    \equiv -\beta^t_2  \cr
  \tilde\alpha^0 a_x^3 + \tilde{w} \tilde\alpha^2
    \quad &\to\quad \beta_x^3
    \equiv \beta^x_3
 }
as $r \to \infty$, and then take mixed derivatives $\partial^2 / \partial\beta^* \partial\beta$ of ${\cal J}$ evaluated at the boundary.  Treating ${\cal J}|_{\rm bdy}$ as the only contribution to the on-shell action means neglecting possible contributions from the horizon, which can be justified along the lines of \cite{Son:2002sd}.  To be precise: we start with a purely infalling solution $a_t^1$, $a_t^2$, and $a_x^3$; then we send the boundary term \eno{JDef} through the gauge transformation \eno{GaugeEquivalents} with the $\tilde\alpha^a$ chosen to satisfy \eno{ChooseatOneBdy}; then we express the whole thing in terms of the $\beta_m^a$'s together with the far-field series expansion coefficients $A_m^{a(s)}$ of the {\it original} infalling solution.  After all this processing, one finds
 \eqn{Jexpress}{
  {\cal J}|_{\rm bdy} =
   {1\over 2 \kappa^2} \begin{pmatrix} \beta_1^{t*} & \beta_2^{t*} & \beta_3^{x*} \end{pmatrix}
    i\omega \tilde{\boldsymbol{\sigma}}
    \begin{pmatrix} \beta^t_{1} \\ \beta^t_{2} \\ \beta^x_{3} \end{pmatrix}
     \,,
 }
with
\eqn{sigmaEntries}{
  \tilde\sigma_{tt}^{11} &=
    \tilde\sigma_{tt}^{22} = {i \over \omega} {\tilde{p}_0 \tilde{p}_1
     \over \tilde{p}_0^2 - \omega^2}  \cr
  \tilde\sigma_{tt}^{12} &= -\tilde\sigma_{tt}^{21} = -{\tilde{p}_1
     \over \tilde{p}_0^2 - \omega^2}  \cr
  \tilde\sigma_{tx}^{13} &=
    -{i \over \omega A_x^{3(0)}} \left( A_t^{1(1)} -
     \tilde{p}_1 {A_t^{1(0)} \tilde{p}_0 +
     i\omega A_t^{2(0)} \over
    \tilde{p}_0^2-\omega^2} \right) \cr
  \tilde\sigma_{tx}^{23} &=
   -{i \over \omega A_x^{3(0)}} \left(A_t^{2(1)} -
     \tilde{p}_1 {A_t^{2(0)} \tilde{p}_0 -
     i\omega A_t^{1(0)} \over
    \tilde{p}_0^2-\omega^2} \right) \cr
  \tilde\sigma_{xt}^{31} &= -{i \over \omega} {\tilde{p}_0 \tilde{W}_1
     \over \tilde{p}_0^2 - \omega^2}  \cr
  \tilde\sigma_{xt}^{32} &= {\tilde{W}_1
     \over \tilde{p}_0^2 - \omega^2}  \cr
  \tilde\sigma_{xx}^{33} &= -{i \over \omega A_x^{3(0)}} \left(
    A_x^{3(1)} + \tilde{W}_1 {A_t^{1(0)} \tilde{p}_0 +
      i\omega A_t^{2(0)} \over \tilde{p}_0^2 - \omega^2} \right) \,.
 }
$\tilde\sigma_{xx}^{33}$ is the same as the one obtained in \eqref{sigmaXXclaim} because the calculation that led to \eqref{sigmaXXclaim} is a special case of the one above.  The far-field limit of the two constraint equations \eqref{FarConstraints} enforce the relations $\tilde\sigma_{tx}^{13} = \tilde\sigma_{xt}^{31}$ and $\tilde\sigma_{tx}^{23} = - \tilde\sigma_{xt}^{32}$, so
 \eqn{FullSigma}{
  \tilde{\boldsymbol{\sigma}} = \begin{pmatrix}
    \displaystyle{{i \over \omega} {\tilde{p}_0 \tilde{p}_1 \over \tilde{p}_0^2 - \omega^2}} &
    \displaystyle{-{\tilde{p}_1 \over \tilde{p}_0^2 - \omega^2}} &
    \displaystyle{-{i \over \omega} {\tilde{p}_0 \tilde{W}_1  \over \tilde{p}_0^2 - \omega^2}} \\[3\jot]
    \displaystyle{{\tilde{p}_1 \over \tilde{p}_0^2 - \omega^2}} &
    \displaystyle{{i \over \omega} {\tilde{p}_0 \tilde{p}_1 \over \tilde{p}_0^2 - \omega^2}} &
    \displaystyle{-{\tilde{W}_1 \over \tilde{p}_0^2 - \omega^2}} \\[3\jot]
    \displaystyle{-{i \over \omega} {\tilde{p}_0 \tilde{W}_1  \over \tilde{p}_0^2 - \omega^2}} &
    \displaystyle{{\tilde{W}_1 \over \tilde{p}_0^2 - \omega^2}} &
    \tilde{\sigma}_{xx}^{33}
  \end{pmatrix} \,,
 }
with $\tilde{\sigma}_{xx}^{33}$ being given in \eqref{sigmaEntries}.  Since $\tilde{p}_0$, $\tilde{p}_1$, and $\tilde{W}_1$ are real, the only contribution to the hermitian (dissipative) part of $\tilde{\boldsymbol{\sigma}}$ comes from $\tilde{\sigma}_{xx}^{33}$.

It is worth noting that given the conductivity matrix $\tilde{\boldsymbol{\sigma}}$ one can compute its inverse $\tilde{\boldsymbol{\rho}} = \tilde{\boldsymbol{\sigma}}^{-1}$.  After imposing \eqref{FarConstraints}, we obtain
 \eqn{GotRho}{
  \tilde{\boldsymbol{\rho}} = \begin{pmatrix}
    \displaystyle{-i \omega {\tilde{p}_0 \over \tilde{p}_1} +
      {\tilde{W}_1^2 \over \tilde{p}_1^2} \tilde{\rho}_{xx}^{33}} &
    \displaystyle{-{\omega^2 \over \tilde{p}_1}} &
    \tilde{\rho}_{xx}^{33} \\[3\jot]
    \displaystyle{{\omega^2 \over \tilde{p}_1}} &
    \displaystyle{-i \omega {\tilde{p}_0 \over \tilde{p}_1}} &
    0 \\[3\jot]
    \tilde{\rho}_{xx}^{33} &
    0 &
    \tilde{\rho}_{xx}^{33}
  \end{pmatrix} \,,
 }
where $\tilde{\rho}_{xx}^{33}$ can be computed from
 \eqn{FoundRho}{
  {1 \over \tilde\rho_{xx}^{33}} = -{i \over \omega}
   {A_x^{3(1)} + A_t^{1(1)} \tilde{W}_1/\tilde{p}_1 \over
    A_x^{3(0)}} \,.
 }
Here $1/\tilde\rho_{33}$ is just the numerical reciprocal of $\tilde\rho_{33}$, not a matrix inverse.  Of the entries of $\tilde{\boldsymbol{\rho}}$, $\tilde\rho_{xx}^{33}$ is the most interesting, because it represents $E_x^3 / J_x^3$ with the constraint that $J_t^1=J_t^2=0$.  Using again the far-field constraints \eqref{FarConstraints}, we obtain
 \eqn{RhoInvMinusSigma}{
  \Im {1\over \tilde\rho_{xx}^{33}} - \Im \tilde\sigma_{xx}^{33} = -{\tilde W_1^2 \tilde p_0 \over \omega \tilde p_1 (\tilde p_0^2 - \omega^2) } \,.
 }
Because $\tilde{W}_1$, $\tilde{p}_0$, and $\tilde{p}_1$ have no $\omega$ dependence, \eno{RhoInvMinusSigma} implies that $\Re 1/\tilde{\rho}^{33}_{xx}$ and $\Re\tilde\sigma^{33}_{xx}$ differ only by a sum of three delta functions at $\omega = 0$ and $\pm\tilde{p}_0$.  The form of \eno{FoundRho} suggests that $\Im 1/\tilde\rho^{33}_{xx}$ doesn't have poles at $\omega = \pm\tilde{p}_0$.  Numerical evaluations confirm this.  So $\Re 1/\tilde\rho^{33}_{xx}$ doesn't have a delta function singularity at $\omega = \pm\tilde{p}_0$.

The singularity structure at $\omega = \pm\tilde{p}_0$ is an unexpected feature of our calculations.  Our heuristic understanding of it hinges on thinking of the boundary limit of $A_t^3$ as a chemical potential for the charge density $J^{t3}$.  The fluctuations that we are tracking in \eno{Jexpress} include small rotations of the $J^{t3}$ charge condensate into the $J^{t1}$ and/or $J^{t2}$ directions.  Assuming that the total magnitude of the condensate remains fixed, this implies a slight decrease of the $J^{t3}$ condensate.  The chemical potential fights against this and tends to pull the condensate back to the $J^{t3}$ direction, similar to gravity's pull on a pendulum.  In the approximations we use, this resonance is undamped, but some kind of corrections---perhaps loop corrections on the gravity side---might damp it and smooth out the delta function singularity into a peak.

\clearpage
\bibliographystyle{ssg}
\bibliography{photon}
\end{document}